\documentclass[twocolumn]{aastex7}
\usepackage{enumerate}
\usepackage{amssymb, amsmath}
\usepackage{natbib}
\usepackage{color}
\usepackage{ulem}
\usepackage{appendix}
\usepackage{hyperref}
\usepackage{relsize}
\usepackage{mathtools}
\usepackage{comment}
\definecolor{red}{rgb}{1.0,0.0,0.0}

\def\surveyname{OASIS}

\definecolor{tccolor}{rgb}{0,0.5,1}

\begin{document}

\title{OASIS Survey Direct Imaging and Astrometric Discovery of HIP 71618 B: \\A Substellar Companion Suitable for the Roman Coronagraph Technology Demonstration\footnote{Based in part on data collected at Subaru Telescope, which is operated by the National Astronomical Observatory of Japan.}}


\author{Mona El Morsy}
\affiliation{Department of Physics and Astronomy, University of Texas at San Antonio, San Antonio, TX 78249, USA}
\email[show]{mona.elmorsy@utsa.edu}
\author[0000-0002-7405-3119]{Thayne Currie}
\affiliation{Department of Physics and Astronomy, University of Texas at San Antonio, San Antonio, TX 78249, USA}
 \affiliation{Subaru Telescope, National Astronomical Observatory of Japan, 
650 North A`oh$\bar{o}$k$\bar{u}$ Place, Hilo, HI  96720, USA}
\email{thayne.currie@utsa.edu}
\author{Brianna Lacy}
\affiliation{Department of Astronomy and Astrophysics, University of California-Santa Cruz, Santa Cruz, CA USA}
\email[]{blacy@ucsc.edu}
\author[0000-0001-8103-5499]{Taylor L. Tobin}
\affiliation{Department of Astronomy, University of Michigan, 1085 S. University, Ann Arbor, MI 48109, USA}
\email{ttobin@umich.edu}
\author[0000-0003-0115-547X]{Qier An}
\affiliation{Department of Physics and Astronomy Johns Hopkins University, Baltimore, MD 21218, USA}
\email[]{qan4@jh.edu}
\author[0000-0002-6845-9702]{Yiting Li}
\affiliation{Department of Astronomy, University of Michigan, 1085 S. University, Ann Arbor, MI 48109, USA}
\email[]{lyiting@umich.edu}
\author{Ziying Gu}
\affiliation{Department of Astronomy, Faculty of Science, The University of Tokyo, 7-3-1 Hongo, Bunkyo-ku, Tokyo 113-0033, Japan}
\email{guziying@g.ecc.u-tokyo.ac.jp}
\author[0000-0002-4677-9182]{Masayuki Kuzuhara}
\affiliation{Astrobiology Center, NINS, 2-21-1 Osawa, Mitaka, Tokyo 181-8588, Japan}
\affiliation{National Astronomical Observatory of Japan, 2-21-1 Osawa, Mitaka, Tokyo 181-8588, Japan}
\email{m.kuzuhara@nao.ac.jp}
\author{Danielle Bovie}
\affiliation{Department of Physics and Astronomy, University of Texas at San Antonio, San Antonio, TX 78249, USA}
\email{danielle.bovie@utsa.edu}
\author{Dillon Peng}
\affiliation{Department of Physics and Astronomy, University of Notre Dame,
Nieuwland Science Hall, Notre Dame, IN 46556, USA}
\email{dpeng@nd.edu}
\author[0000-0001-6305-7272]{Jeffrey Chilcote}
\affiliation{Department of Physics and Astronomy, University of Notre Dame,
Nieuwland Science Hall, Notre Dame, IN 46556, USA}
\email{jchilcot@nd.edu}

\author[]{Olivier Guyon}
 \affiliation{Subaru Telescope, National Astronomical Observatory of Japan,
650 North A`oh$\bar{o}$k$\bar{u}$ Place, Hilo, HI  96720, USA}
\affiliation{Astrobiology Center, 2-21-1 Osawa, Mitaka, Tokyo 181-8588, Japan}
\email[]{guyon@naoj.org}

\author[0000-0001-6341-310X]{Miles Lucas}
\affiliation{Steward Observatory, University of Arizona, Tucson, AZ 87521, USA}
 \affiliation{Subaru Telescope, National Astronomical Observatory of Japan,
650 North A`oh$\bar{o}$k$\bar{u}$ Place, Hilo, HI  96720, USA}
\email[]{mileslucas@arizona.edu}

\author{Timothy D Brandt}
\affiliation{Space Telescope Science Institute, Baltimore, MD, USA}
\email[]{tbrandt@stsci.edu }
\author{Robert de Rosa}
\email[]{rderosa@eso.org}
\affiliation{European Southern Observatory, Alonso de Córdova 3107, Vitacura, Santiago, Chile}
\author{Tyler D Groff}
\affiliation{NASA-Goddard Space Flight Center, Greenbelt, MD USA}
\email[]{tyler.d.groff@nasa.gov}
\author{Markus Janson}
\affiliation{Department of Astronomy, Stockholm University, AlbaNova University Center, Stockholm, 10691, Sweden}
\email[]{markus.janson@astro.su.se}
\author{N. J. Kasdin}
\affiliation{Department of Mechanical and Aerospace Engineering Princeton University, Princeton, New Jersey 08544, USA}
\email[]{jkasdin@princeton.edu}
\author[0000-0002-3047-1845]{Julien Lozi}
 \affiliation{Subaru Telescope, National Astronomical Observatory of Japan,
650 North A`oh$\bar{o}$k$\bar{u}$ Place, Hilo, HI  96720, USA}
\affiliation{Astrobiology Center, 2-21-1 Osawa, Mitaka, Tokyo 181-8588, Japan}
\email[]{lozi@naoj.org}
\author{Christian Marois}
\affiliation{National Research Council, Herzberg Astronomy and Astrophysics, Victoria, BC, Canada}
\email{christian.marois@nrc-cnrc.gc.ca}
\author{Bertrand Mennesson}
\affiliation{Jet Propulsion Laboratory, California Institute of Technology, Pasadena, California, United States}
\email[]{Bertrand.Mennesson@jpl.nasa.gov}
\author[0000-0001-6177-1333]{Naoshi Murakami}
\affiliation{Astrobiology Center, NINS, 2-21-1 Osawa, Mitaka, Tokyo 181-8588, Japan}
\affiliation{National Astronomical Observatory of Japan, 2-21-1 Osawa, Mitaka, Tokyo 181-8588, Japan}
\affiliation{The Graduate University for Advanced Studies, SOKENDAI, 2-21-1 Osawa, Mitaka, Tokyo 181-8588, Japan}
\affiliation{Faculty of Engineering, Hokkaido University, Kita 13, Nishi 8, Kita-ku, Sapporo, Hokkaido 060-8628, Japan}
\email[]{naoshi.murakami@nao.ac.jp}

\author{Eric Nielsen}
\affiliation{Department of Astronomy, New Mexico State University, P. O. Box 30001, MSC 4500, Las Cruces, NM 88003, USA}
\email[]{nielsen@nmsu.edu}

\author{Sabina Sagynbayeva}
\affiliation{ Department of Physics and Astronomy, Stony Brook University, Stony Brook, NY 11794, USA  }
\email[]{sabina.sagynbayeva@stonybrook.edu}
\author[0000-0002-9372-5056]{Nour Skaf}
\affiliation{Institute for Astronomy, University of Hawai’i at Manoa, Hilo, HI 96720-2700, US}
\email[]{nskaf@hawaii.edu}

\author{William Thompson}
\affiliation{National Research Council, Herzberg Astronomy and Astrophysics, Victoria, BC, Canada}
\email[]{William.Thompson@nrc-cnrc.gc.ca}
\author[0000-0002-6510-0681]{Motohide Tamura}
\affiliation{Astrobiology Center, 2-21-1 Osawa, Mitaka, Tokyo 181-8588, Japan}
\affiliation{National Astronomical Observatory of Japan, 2-21-1 Osawa, Mitaka, Tokyo 181-8588, Japan}
\affiliation{Department of Astronomy, Faculty of Science, The University of Tokyo, 7-3-1 Hongo, Bunkyo-ku, Tokyo 113-0033, Japan}
\email[]{motohide.tamura@astron.s.u-tokyo.ac.jp}

\author[0000-0002-6879-3030]{Taichi Uyama}
\affiliation{Department of Physics and Astronomy, California State University Northridge, 18111 Nordhoff Street, Northridge, CA 91330, USA}
\email[]{taichi.uyama.astro@gmail.com}

\author[0000-0002-5903-8316]{Alice Zurlo}
\affiliation{Instituto de Estudios Astrof\'isicos, Facultad de Ingenier\'ia y Ciencias, Universidad Diego Portales, Av. Ej\'ercito Libertador 441, Santiago, Chile}
\affiliation{Millennium Nucleus on Young Exoplanets and their Moons (YEMS)}
\email[]{alice.zurlo@mail.udp.cl}

\author{Vincent Deo}
 \affiliation{Subaru Telescope, National Astronomical Observatory of Japan,
650 North A`oh$\bar{o}$k$\bar{u}$ Place, Hilo, HI  96720, USA}
\email[]{vdeo@naoj.org}
\author{Sebastien Vievard}
 \affiliation{Subaru Telescope, National Astronomical Observatory of Japan,
650 North A`oh$\bar{o}$k$\bar{u}$ Place, Hilo, HI  96720, USA}
\email[]{vievard@naoj.org}

\shortauthors{El Morsy et al.}
\begin{abstract}
We present the OASIS survey program discovery of a substellar companion orbiting the young A1V star HIP 71618, detected using precision astrometry from \textit{Gaia and Hipparcos} and high-contrast imaging with SCExAO/CHARIS and Keck/NIRC2.   Atmospheric modeling favors a spectral type of M5--M8 and a temperature of $\sim$2700 $\pm$ 100 $K$.
Dynamical modeling constrains HIP 71618 B's mass to be ${60}_{-21}^{+27}$ $M_{\rm Jup}$ or ${65}_{-29}^{+54}$ $M_{\rm Jup}$, depending on the adopted companion mass prior.  It has a nearly-edge-on, 11 au orbit with a high eccentricity.
HIP 71618 B will be located within Roman Coronagraph's dark hole region during the instrument's technological demonstration phase.  A high signal-to-noise-ratio detection of HIP 71618 B at 575 nm would demonstrate a 5-$\sigma$ contrast of 10$^{-7}$ or better.  The system is also located within or very close to Roman's Continuous Viewing Zone -- near multiple candidate reference stars for dark-hole digging -- and its primary is bright ($V$ $\approx$ 5).  
The suitability of HIP 71618 as one potential Roman Coronagraph target for demonstrating the instrument's core requirement (TTR5) should motivate the timely, deep vetting of candidate reference stars.

\end{abstract}

\section{Introduction}

High-contrast imaging observations over the past $\sim$15-20 years have revealed about two-dozen directly imaged superjovian planets \citep[][]{Marois2008a,Lagrange2010,Currie2023b}.   Many recent planet imaging campaigns target stars based on direct, dynamical evidence for an imageable companion from Hipparcos and Gaia precision astrometry \citep[e.g.][]{Currie2023a,Franson2023,deRosa2023,Mesa2023,ElMorsy2024a}.

While still in their infancy, these surveys have already yielded the discovery of numerous substellar companions from ground-based adaptive optics (AO) assisted near-infrared (IR) data, including three superjovian planets (HIP 99770 b, AF Lep b, and HIP 54515 b) \citep{Currie2020a,Kuzuhara2022,Currie2023a,deRosa2023,Mesa2023,Franson2023,Currie2025a}.  Compared to traditional, so-called ``blind" or unbiased surveys \citep[e.g.][]{Nielsen2019}, these campaigns appear to yield a higher rate of discoveries \citep{ElMorsy2024a}.   Jointly modeling direct imaging and astrometric data can directly constrain the companions' masses and improve orbital parameters estimates compared to direct imaging data alone  \citep[e.g.][]{Brandt2019,Brandt2021c}.

The companions found from these direct imaging and astrometry programs are well-suited for detailed follow-up characterization and strategically important observations for NASA missions.   Near-IR spectroscopic follow-up data better constrain companion chemistries and gravities \citep[e.g.][]{Bovie2025,ElMorsy2024b}, including at moderate to high resolutions capable of constraining molecular abundances \citep{Balmer2023,Balmer2025,Zhang2024,Winterhalder2025}.  Thermal IR follow-up (e.g. with JWST/NIRCam) can clarify atmospheric chemistry \citep{Franson2024}.  Self-luminous substellar companions may have optical contrasts making them suitable for the critical Roman Coronagraph Instrument technology demonstration \citep{LacyBurrows2020}.

In this Letter, we report the discovery of a brown dwarf orbiting $\sim$ 11 au from the A-type star HIP 71618 from the Observing Accelerators with SCExAO Imaging Survey \citep[OASIS; PI: T. Currie, Co-PI: M. Kuzuhara][]{ElMorsy2024a}.  The discovery combines precision astrometry from the Hipparcos-Gaia Catalogue of Accelerations \citep{Brandt2021b}, near-IR high-contrast  direct imaging and spectroscopy from SCExAO/CHARIS, and thermal IR imaging from Keck/NIRC2.   In addition to being the second OASIS discovery, HIP 71618 B is the first substellar companion demonstrably suitable for the Roman Coronagraph Instrument's Technology Demonstration.

\begin{figure}
\centering
   \includegraphics[width=0.5\textwidth]{gaia_color_diagram_hip71618.pdf}
   \vspace{-0.2in}
   \label{fig:cmd} 
   \caption{The \textit{Gaia} color-magnitude diagram for the Pleiades (grey) and Hyades (dark red) compared to PARSEC isochrones for 112 Myr and 750 Myr, young stars with interferometrically measured ages $\lesssim$ 150 Myr, stars in the 400 Myr-old Ursa Majoris Moving group, and other young stars with imaged substellar companions \citep{Jones2015,Jones2016,Jones2016PhD,Currie2023a,Tobin2024}.}
\end{figure}

\section{Properties of the HIP 71618 System\label{sec:system_properties}} 
HIP 71618 (33 Boo, HD 129002) is a bright, nearby ($G$ = 5.38, d = 51.92 $pc$) A1V  \citep{GAIA2023}.  Banyan-$\Sigma$ \citep{Gagne2018} does not identify HIP 71618 as a high-confidence member of any known moving group (p(field )$\sim$ 98\%, p(AB Dor) $\sim$ 2\%) \citep{Gagne2018}.   However, prior analyses based on color-magnitude diagrams find that the star is young, albeit with a very large dispersion of estimated ages \citep[i.e. 8-220 Myr;][]{deRosa2014,David2015}.   

Table \ref{tab:prop_c} lists the star's properties, derived either from the literature or from our analysis below.  We use empirical comparisons to stars with well-constrained ages to provide a context for HIP 71618's likely age.  Its \textit{Gaia} color-magnitude diagram position is consistent with those of stars in the Pleiades ($\sim$115 Myr),  $\kappa$ And ($t$ $\sim$ 47 Myr), $\beta$ Pic ($t$ $\sim$ 24 Myr) and other young field stars with interferometrically-determined of ages less than 150 Myr, while it lies closer to the Pleiades sequence than HIP 99770 ($t$ $\sim$ 115-200 Myr) and far closer than stars in the 414 Myr-old Ursa Majoris moving group \citep[][Figure \ref{fig:cmd}]{Jones2016,Currie2023a}.   Ages less than 20 Myr are arguably disfavored since HIP 71618 is not a known member of any known cluster.  We thus nominally adopt an age of 115$^{+95}_{-95}$ Myr, where the lower bounds are set by its lack of membership in a known moving group and upper bounds set by its youth relative to HIP 99770.   From the relationship mapping between stellar mass, spectral type, and bolometric correction for a star on the main sequence \citep{Pecaut2013}, we favor a stellar mass of 2.05 $M_{\rm \odot}$, temperature of 9300 $K$ and luminosity of log($L_{\rm bol}$/$L_{\odot}$) = 1.327.

Previous AO observations did not identify a candidate imaged companion  \citep{deRosa2011,deRosa2014}, although they identify a candidate ROSAT x-ray source coincident with the star.  The star was not observed before with extreme AO platforms (e.g., the Gemini Planet Imager).

The {Hipparcos}-{Gaia} Catalog of Accelerations identifies HIP 71618 as having a proper motion anomaly of $\chi^{2}$ = 9.42 or $\sim$2.61$\sigma$ for two degrees of freedom \citep{Brandt2021b}.  Similarly, the \textit{Gaia eDR3} astrometric catalogue from \citet{Kervella2022} identifies a $\sim$2.7-$\sigma$ significant astrometric acceleration\footnote{The Gaia eDR3 catalogue identifies HIP 71618 as having a moderately-high Renormalized Unit Weight Error that may also indicate evidence for a companion, but this property did not originally motivate our targeting of the star.}.  The acceleration derived from HGCA is consistent with a substellar companion at 0\farcs{}5 (0\farcs{}25) with a minimum mass of $\approx$ 60 $M_{\rm Jup}$ (22 $M_{\rm Jup}$).  Thus, we targeted this star with \surveyname\ survey \citep[PI T. Currie;][]{ElMorsy2024a}.

\begin{deluxetable}{lcc}
\setlength{\tabcolsep}{0pt}
\tablecaption{\label{tab:prop_c}Properties of HIP 71618}
\tablehead{\colhead{Property} & \colhead{Value} & \colhead{Refs}}
\startdata
$\alpha_{2000.0}$ & 14:38:50.2 & 1\\
$\delta_{2000.0}$ & +44:24:16.21 & 1\\
Distance (pc) & 58.19 $\pm$ 0.51 & 1\\
$\pi$ (mas) & 17.1844 $\pm$ 0.1516 & 1\\
$\mu_{\alpha}$, $\mu_{\delta}$ H-G$^{a}$ (mas/yr) & -68.287 $\pm$ 0.015, -18.122 $\pm$ 0.015 & 2\\
$\mu_{\alpha}$, $\mu_{\delta}$ G$^{b}$ (mas/yr) & -67.847 $\pm$ 0.174, -18.424 $\pm$ 0.190 & 2\\
$\mu_{\alpha}$, $\mu_{\delta}$ H$^{c}$ (mas/yr) & -68.218 $\pm$ 0.506, -18.700 $\pm$ 0.440 & 2\\
Proper Motion Anomaly & $\chi^{2}$=9.421, 2.612$\sigma$ & 2 \\
Mass ($\mathrm{M_{\odot}}$) & 2.05$^{+0.13}_{-0.07}$ & 3\\
Spectral Type & A1V & 4\\
$T_{\rm eff}$ (K) & 9300$^{+400}_{-500}$ & 3\\
$\log(L_{\mathrm{bol}} / L_\odot)$ (dex) & 1.327 $\pm$ 0.029 & 3\\
Angular Diameter (mas) & 0.291 $\pm$ 0.0151 & 5\\
$V$ (mag) & 5.387 $\pm$ 0.009 & 6
\enddata
\tablecomments{References: (1) - \citet{GAIA2023}, (2) - \citet{Brandt2021b}, (3) - \citet{Pecaut2013}/this work, (4) - \citet{Wenger2000}, (5) - \citet{Kervella2022}, (6) - \citet{Tycho2}. a,b,c - proper motion measurements drawn from the Hipparcos-Gaia scaled positional distance (H-G), Gaia eDR3 (G), and Hipparcos (H) in \citet{Brandt2021b}.}
\end{deluxetable}

\begin{deluxetable*}{lllllllll}[ht!]
     \tablewidth{0pt}
    \tablecaption{HIP 71618 Observing Log\label{obslog}}
    \tablehead{\colhead{UT Date} & \colhead{Instrument} &  \colhead{Seeing$^{c}$ (\arcsec{})} &{Filter} & \colhead{$\lambda$ ($\mu m$)$^{a}$} 
    & \colhead{$t_{\rm exp}$} & \colhead{$N_{\rm exp}$} & \colhead{$\Delta$PA ($^{o}$)} & SNR$^{d}$}
    \startdata
    20240220 & AO188+SCExAO/CHARIS &  0.7 & $JHK$ & 1.15-2.37& 25.08 & 88 & 41.1  & 84.1\\
    20240223 & Keck/NIRC2 &  0.7 & $L_{\rm p}$ & 3.78& 30 &  85& 27.19 &  9.1\\
    20240617 & Keck/NIRC2 &  0.7 & $L_{\rm p}$ & 3.78& 30 & 43 & 50.59 & 15.8 \\
    20250314 & AO3K+SCExAO/CHARIS & 1-3  & $JHK$ & 1.15-2.37 & 20.65 & 257 & 52.38  & 27.6\\
    20250512& AO3K+SCExAO/CHARIS & 0.6  & $JHK$ & 1.15-2.37 & 25.08 & 18 & 5.06 & 26.6$^{b}$
    \enddata
    \tablecomments{
    a) 
   For CHARIS data, the $\lambda_{\rm \mu m}$ column refers to the wavelength range.  For NIRC2 imaging data, it refers to the central wavelength. 
    b)   Reduced with SDI only: extracted spectrum is less reliable.}
    \label{obslog_hip71618}
    \end{deluxetable*}
\section{Data\label{sec:data}}
\subsection{Observation and Data Reduction }

We observed HIP 71618 in five epochs between 2024 February and 2025 Mayusing the Subaru and Keck telescopes located at Maunakea (see Table~\ref{obslog_hip71618}). At Subaru, the facility AO systems -- AO188 in 2024 and AO3K in 2025 -- performed the initial correction of atmospheric turbulence \citep{Minowa2010,Lozi2022}. 
SCExAO then performed higher-order wavefront error correction and sent the light to the CHARIS integral field spectrograph (IFS) \citep{Groff2016}.  
For Keck, we obtained NIRC2 imaging in the $L_{\rm p}$ broadband filter ($\lambda_{o}$ = 3.78 $\mu m$) behind Keck II's facility Shack-Hartmann AO system.   

For the 2025 March SCExAO/CHARIS observations, the seeing was extremely poor (up to 3\arcsec{} seeing), yet AO3K+SCExAO was able to sustain an AO correction and usually produce a dark hole.  The 2024 June Keck data were obtained through highly variable clouds.  Otherwise, seeing conditions were slightly below average for Maunakea and photometric.    All data were acquired in pupil-tracking mode, allowing the sky to rotate relative to the detector and thus enabling angular differential imaging \citep[ADI;][]{Marois2006}, with integration times of 7.5--88 minutes.

SCExAO/CHARIS data use a Lyot coronagraph with a 0\farcs{}113 focal plane mask to suppress starlight.  All CHARIS data were recorded in low spectral resolution/``broadband" mode covering the JHK passbands simultaneously ($\lambda$ = 1.15--2.39 $\mu m$, $R$ $\sim$ 18).   For image registration and spectrophotometric calibration, we modulated the SCExAO deformable mirror (DM) to produce satellite spots $\sim$16 $\lambda$/D from the star \citep{Jovanovic2015-astrogrids}.

We extracted the CHARIS data cubes within the Automated Data Extraction, Processing, and Tracking System (ADEPTS) \citep[2024 data,][]{Tobin2020,Tobin2022} or independently with identical settings (2025 data), following \citet{Brandt2017}. We further processed data with the CHARIS data processing pipeline \citep{Currie2020b} as in previous work \citep{Currie2023a,Tobin2024}.  Key reduction steps include sky subtraction, image registration, spectrophotometric calibration, PSF subtraction, and spectral extraction. For spectrophotometric calibration, we used a Kurucz model atmosphere \citep{Castelli2003} appropriate for an A1V star.

For the February 2024 and May 2025 data sets, HIP 71618 B was visible in the raw, quicklook CHARIS images produced in real time.   
Thus, our PSF subtraction approach used rather conservative settings with the adaptive locally optimized combination of images (A-LOCI) algorithm \citep{Currie2012,Currie2015} minimizing signal loss, applying a pixel mask over the subtraction zone \citep{Marois2010,Currie2012} and constructing a reference PSF to minimize residuals within the surrounding optimization region \citep{Lafreniere2007}.  For the 2024 February and 2025 March data, we performed PSF subtraction using ADI only, using rotation gaps of $\delta$ = 1.5 and 0.75, respectively \citep[see][]{Lafreniere2007}.  The 2025 May data covered a very small parallactic angle motion, so we performed these data using spectral differential imaging \citep[SDI;][]{Marois2000,SparksFord2002} with a radial scaling gap of $\delta$ = 1. 

We reduced the NIRC2 data using the ADI-based pipeline from \citet{Currie2011}, following reduction steps as described in \citet{ElMorsy2024b}.  
Because HIP 71618 B is a more challenging detection for NIRC2, our PSF subtraction settings are slightly more aggressive, consistently using a smaller rotation gap threshold of $\delta$ = 0.75 and no pixel mask over the subtraction zone.

\subsection{Detections, Spectra, and Astrometry}
\begin{figure*}[ht!]
    \includegraphics[width=0.33\textwidth]{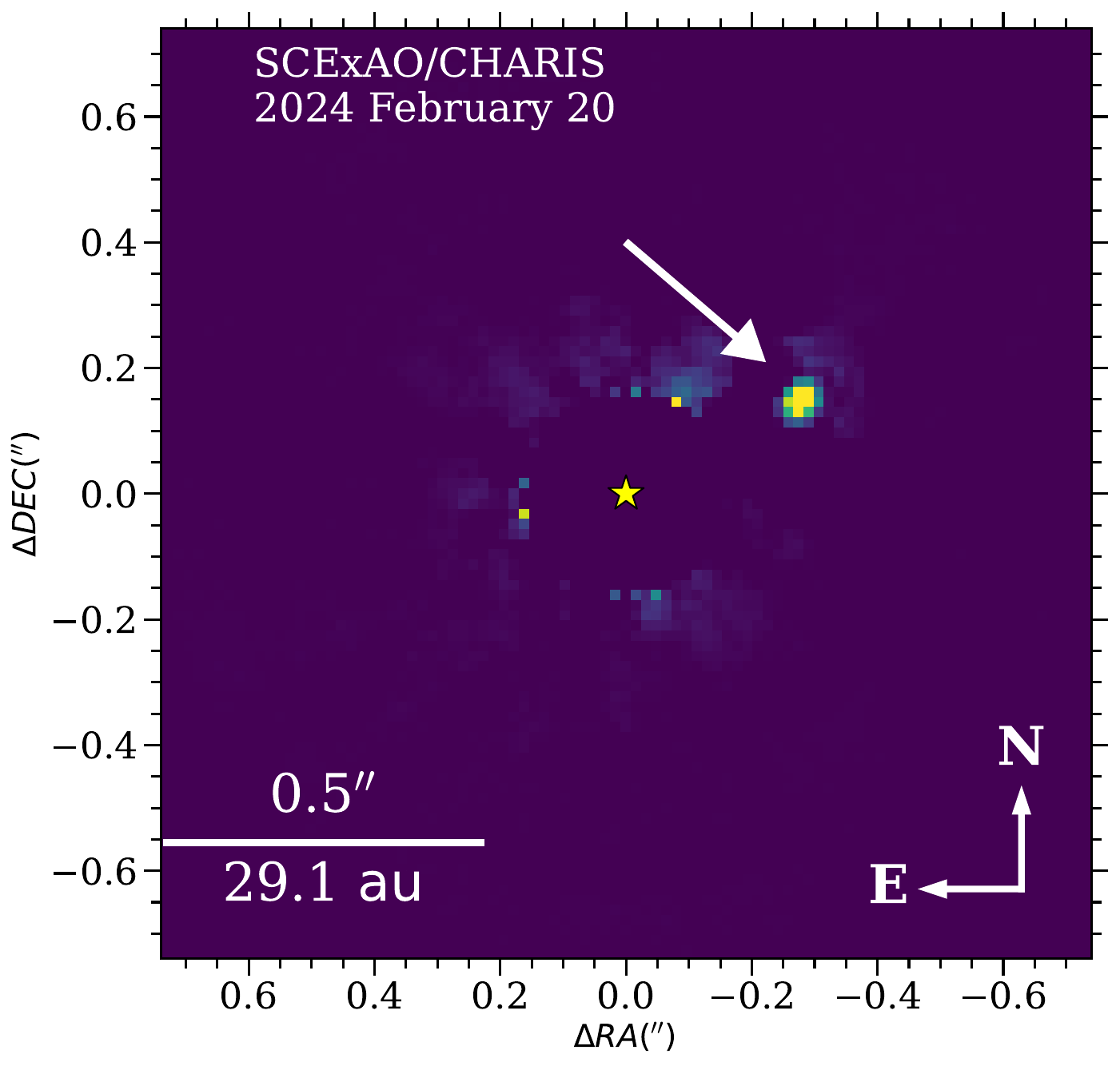}%
    \includegraphics[width=0.33\textwidth]{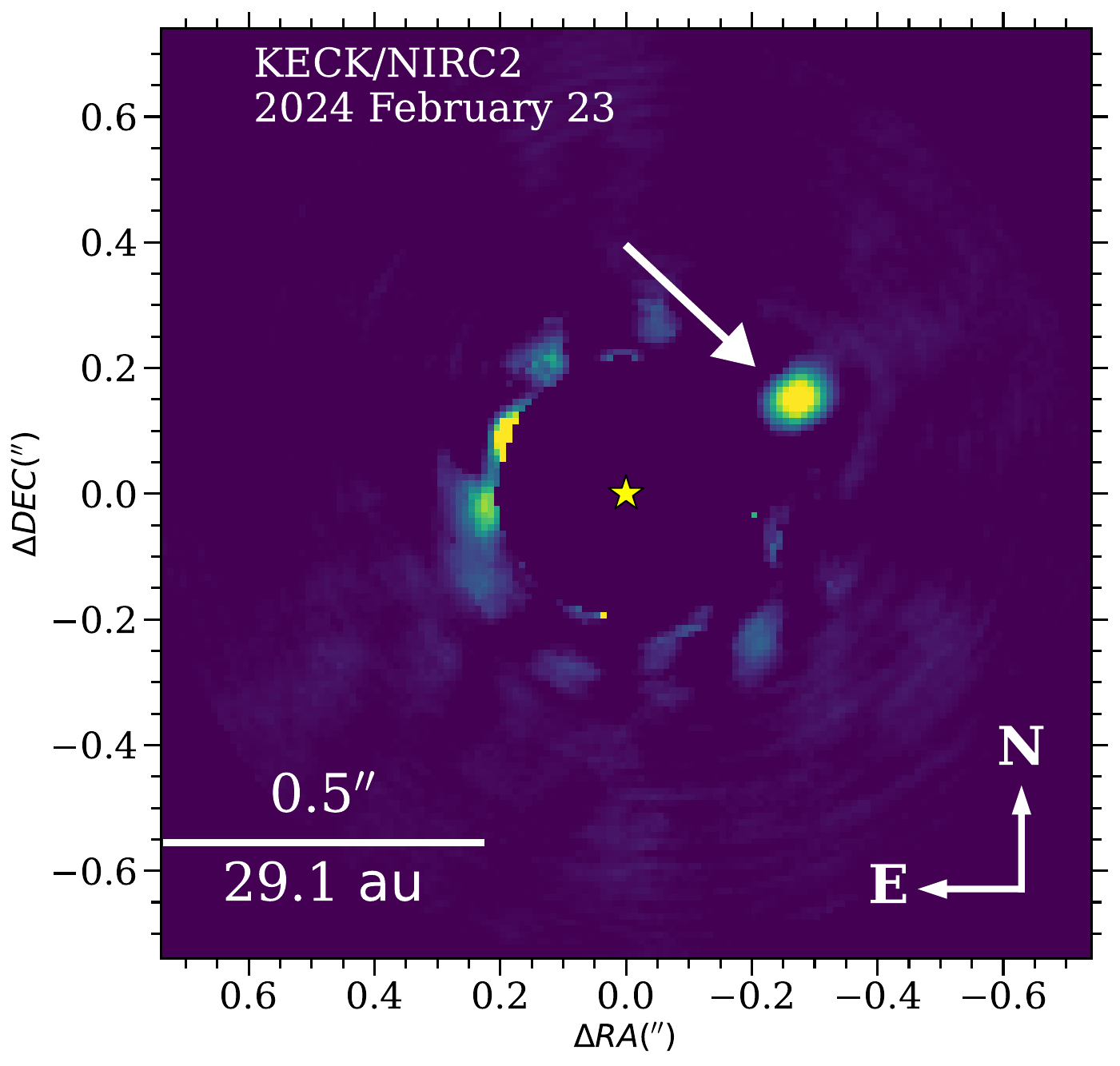}
    \includegraphics[width=0.33\textwidth]{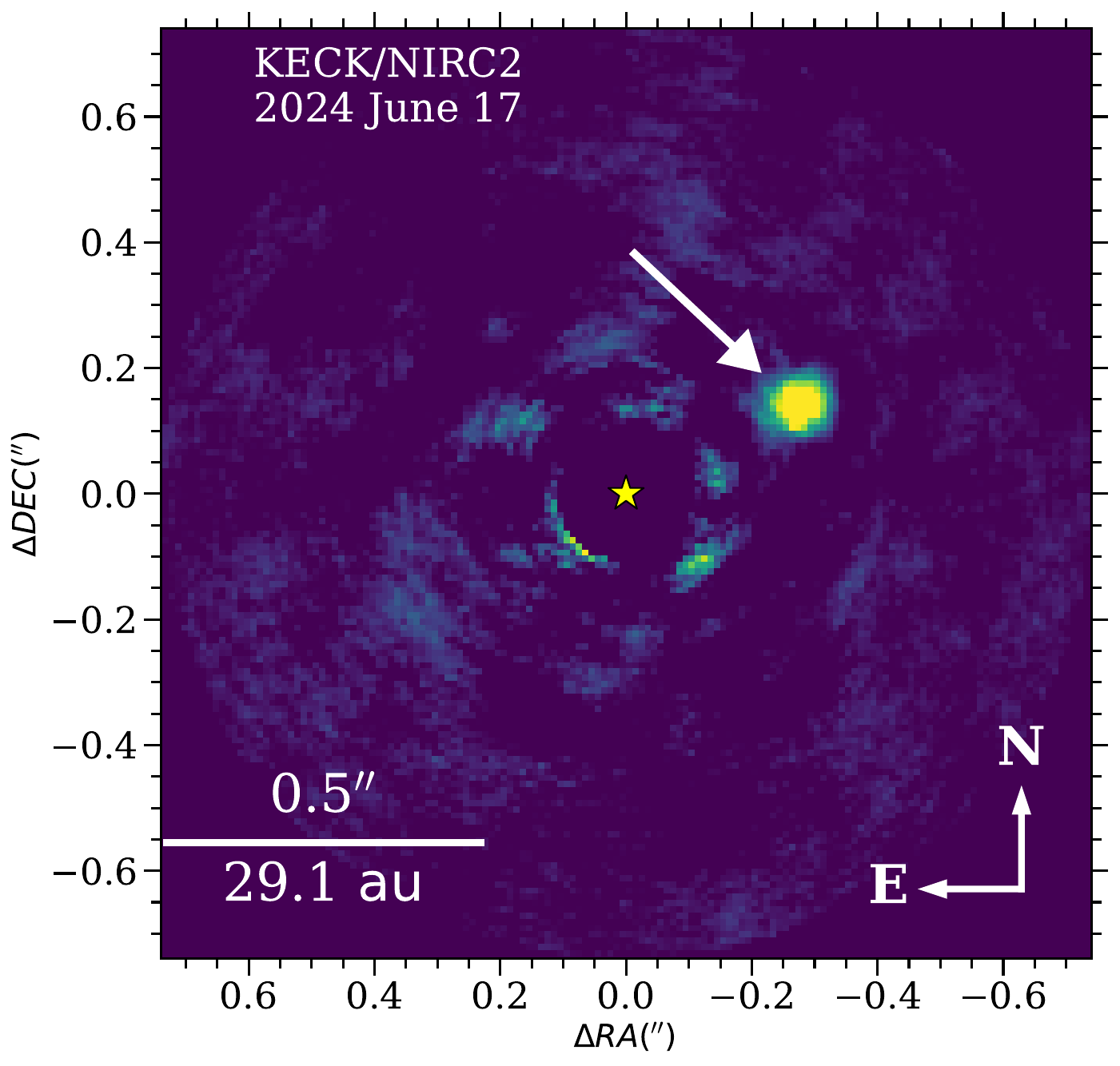}
    \includegraphics[width=0.33\textwidth]{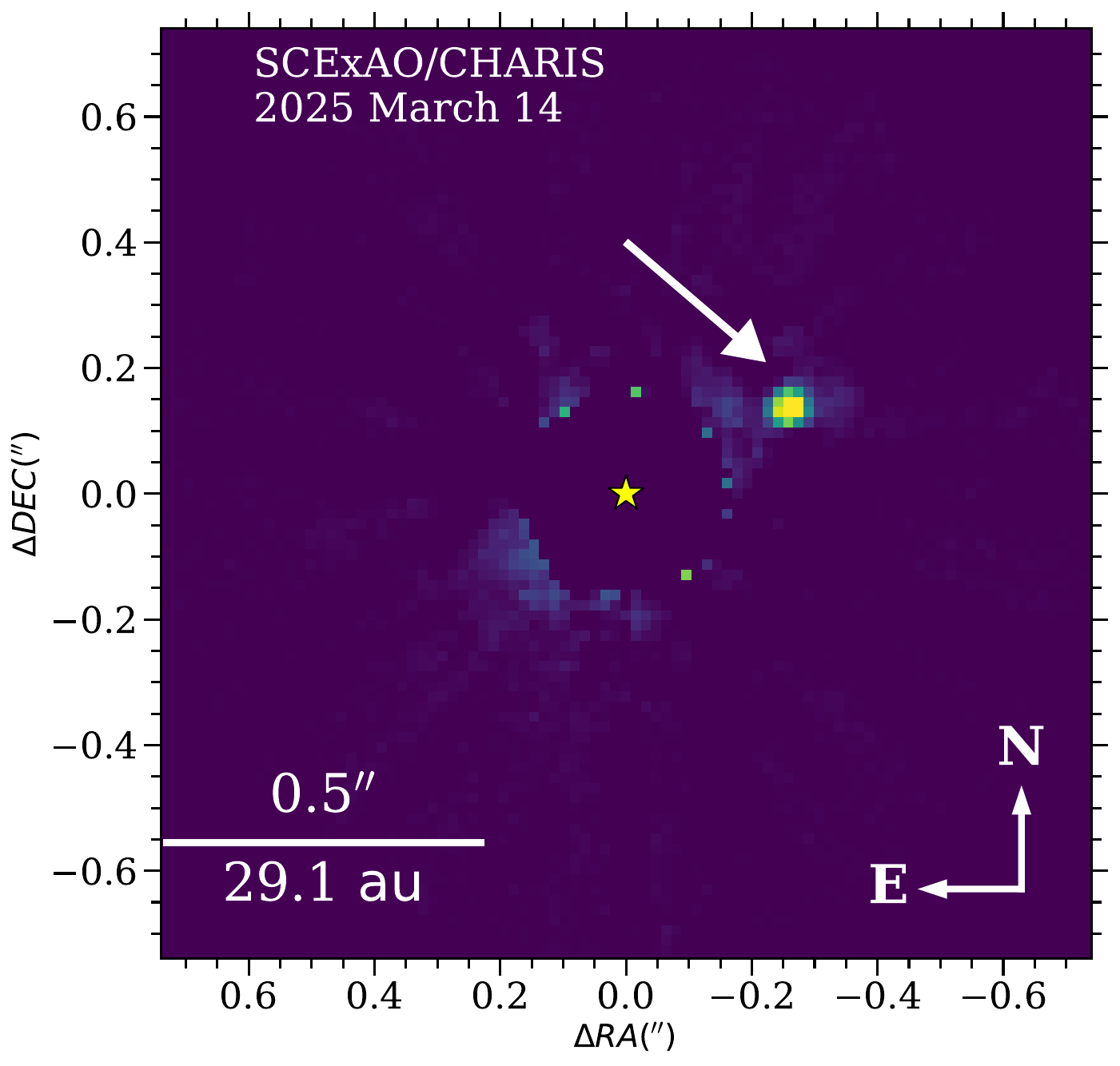}%
    \includegraphics[width=0.33\textwidth]{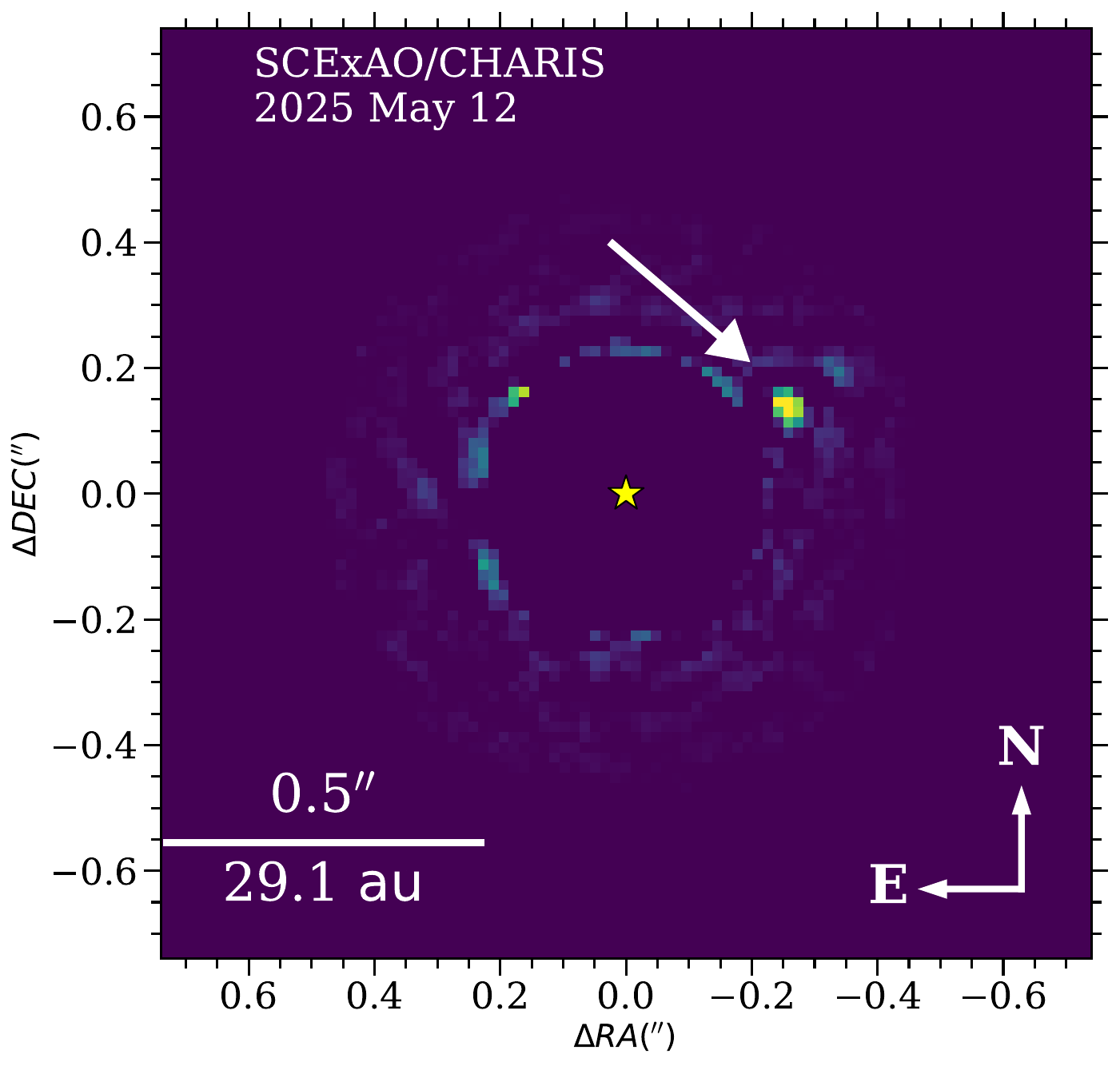}
  \caption{
  Wavelength-collapsed images of HIP 71618 B with SCExAO/CHARIS in broadband near-infrared light (1.15--2.37 $\mu m$) and with Keck/NIRC2 in the $L_{\rm p}$ filter (3.78 $\mu m$).  The white arrow denotes the position of HIP 71618 B in each panel.
  }
  \label{fig:images}
\end{figure*}

Figure \ref{fig:images} shows the detection of HIP 71618 B at roughly the 2 o'clock position $\approx$0\farcs{}3 from the star.  Adopting the standard definition for signal-to-noise ratio (SNR) \citep[e.g.][]{Currie2011} -- corrected for finite sample sizes as in \citet{Mawet2014} -- HIP 71618 B is visible at SNR = 26-81 in the CHARIS data sets and SNR = 9.1--15.8 with NIRC2.  Comparing the earliest and latest CHARIS images suggests a slight decrease in HIP 71618 B's projected separation.

To correct for spectrophotometric and astrometric biases in the SCExAO/CHARIS data introduced by PSF subtraction, we apply forward modeling described in \citet{Currie2018}.  
To empirically estimate astrometric uncertainties, we adopt the approach described in \citet{Kuzuhara2022}: we inject a forward-model PSF at the same angular separation as the object but we vary the azimuthal angle. We then compare the input and recovered centroid positions in pixels for an empirical estimate of the centroiding uncertainty, which we then combine quadratically with the stellar centroid uncertainty.  Finally, we convert our astrometry to polar coordinates: the final astrometric error budget considers the centroiding uncertainty, the uncertainty in the star's position (set at 0.25 pixels), and the uncertainty in the north position angle and pixel scale.

Our CHARIS spectrum extracted from the 2024 February epoch is listed in Table \ref{tab:spectrum_snr}; relative astrometric positions and uncertainties are listed in Table \ref{astrom}, where the table caption lists our adopted astrometric calibration.   
The spectrum extracted from March 2025 is noisier but otherwise agrees with the February 2024 spectrum within 1-$\sigma$ errors. 
Calibrating HIP 71618 B's spectrum from the May 2025 data is more challenging because it was processed with SDI \citep[see][]{Pueyo2012,Pueyo2016}.  Thus, we focus on the 2024 February CHARIS data and NIRC2 data for atmospheric analysis.

We calculate the spectral covariance matrix following the parameterization in \citet{GrecoBrandt2016}. 
We find a best-fitting model of $A_{\rho} = 0.2164$, $A_{\lambda} = 0.5745$, $A_{\sigma} = 0.2091$, ${\sigma}_{\rho} = 0.4309$, and ${\sigma}_{\lambda} = 0.4585$ \citep[see ][for details]{GrecoBrandt2016}.   Thus, spectrally correlated noise dominates the residual speckle properties.

 From CHARIS, we then estimate photometry of $J, H, K_{\rm s}$ = 13.58 $\pm$ 0.06, 13.10 $\pm$ 0.05, and 12.78 $\pm$ 0.09.   For NIRC2, variable clouds preclude a precise photometric calibration from the 2024 June epoch, while our 2024 February data yield $m_{\rm L_{\rm p}}$ = 12.27 $\pm$ 0.19.

\begin{deluxetable}{llll}[!h]
\label{tab:spectrum_snr}
\vspace{-.2in}
     \tablewidth{0pt}
    \tablecaption{HIP 71618 B Spectrum Extracted from February 2024 CHARIS Data}
    \tablehead{\colhead{Wavelength ($\mu$m)} & \colhead{$F_{\rm \nu}$ (mJy)} &  \colhead{$\sigma$~$F_{\rm \nu}$ (mJy)} & \colhead{SNR}}
    \startdata
1.160 & 5.441 & 0.750 & 28.7 \\
1.200 & 5.551 & 0.486 & 25.9 \\
1.241 & 5.268 & 0.284 & 31.1 \\
1.284 & 6.124 & 0.306 & 41.1 \\
1.329 & 5.498 & 0.272 & 41.2 \\
1.375 & 3.761 & 0.194 & 32.1 \\
1.422 & 4.300 & 0.212 & 33.1 \\
1.471 & 4.648 & 0.201 & 45.5 \\
1.522 & 5.067 & 0.240 & 34.1 \\
1.575 & 5.663 & 0.277 & 35.1 \\
1.630 & 6.429 & 0.347 & 27.1 \\
1.686 & 6.632 & 0.332 & 33.0 \\
1.744 & 6.091 & 0.351 & 24.2 \\
1.805 & 4.455 & 0.411 & 23.6 \\
1.867 & 4.962 & 0.294 & 24.3 \\
1.932 & 5.110 & 0.273 & 35.1 \\
1.999 & 4.501 & 0.270 & 35.0 \\
2.068 & 5.078 & 0.383 & 25.8 \\
2.139 & 5.333 & 0.354 & 29.6 \\
2.213 & 5.290 & 0.518 & 20.1 \\
2.290 & 5.347 & 0.505 & 20.5 \\
2.369 & 3.277 & 0.936 & 18.8 \\
\enddata

\end{deluxetable}

\begin{deluxetable*}{lllll}
 \label{astrom}
     \tablewidth{0pt}
     \tabletypesize{\scriptsize}
      \setlength{\tabcolsep}{1pt}
     \tablecaption{HIP 71618 B Astrometry}
      \tablehead{\colhead{UT Date} & \colhead{Instrument}  & \colhead{Filter} & \colhead{$\rho$ $\arcsec{}$} & \colhead{PA (deg)}}
     \startdata
    20240220  & SCExAO/CHARIS & $JHK$ & 0.311 $\pm$ 0.003 & 297.896 $\pm$ 0.601\\
    20240223  & Keck/NIRC2 & $L_{\rm p}$ & 0.304 $\pm$ 0.009 & 298.282 $\pm$ 0.380\\
    20240624  & Keck/NIRC2 & $L_{\rm p}$ & 0.303 $\pm$ 0.004 & 297.038 $\pm$ 0.377\\
    20250314 & SCExAO/CHARIS & $JHK$ & 0.295 $\pm$ 0.004 & 297.634 $\pm$ 0.625\\
    20250512 & SCExAO/CHARIS & $JHK$ & 0.289 $\pm$ 0.006 & 297.343 $\pm$ 0.647
    \enddata
    \tablecomments{We adopt a pixel scale and north position angle offset of 16.10 $\pm$ 0.04 mas pixel$^{-1}$, -2.03 $\pm$ 0.27$^{o}$ for CHARIS and 9.971 $\pm$ 0.004 mas pixel$^{-1}$, 0.262 $\pm$ 0.020$^{o}$ for NIRC2 \citep{Currie2025a,Service2016}.   Using the older CHARIS astrometric calibration from \citet{Currie2018,Currie2022} instead yields almost identical results for our dynamical modeling.}
\end{deluxetable*}

\section{Atmosphere Analysis\label{sec:spec_analysis}}

\subsection{Empirical Constraints on HIP 71618 B's Atmosphere}
We compare the JHK band spectra of HIP 71618 B from the February 2024 SCExAO/CHARIS observation with  MLT substellar objects from the Montreal Spectral Library \citep{Gagne2015}, using Equation 7 from \citet{Currie2018} to quantify the goodness-of-fit, considering both the spectroscopic errors and the spectral covariance.

HIP 71618 B is best fit by the low-gravity M6 dwarf 2MASSJ0318-3708(Figure \ref{fig:fit_btsettl}; $\chi^{2}_{\rm \nu}$ $\sim$ 1.4).   While low-gravity M6 dwarfs do not have effective temperatures listed in \citet{Pecaut2013}, their older field counterparts have temperatures of $\sim$ 2800 $K$: young, low-gravity mid M dwarfs (i.e. M4--M5) have temperatures $\approx$ 40--170 $K$ cooler.  
The full range of spectral types yielding comparable fit qualities ($\chi^{2}_{\rm \nu}$ $\sim$ 1.4--2) is dominated by M6--M8 intermediate- and low-gravity dwarfs ($<$ 2570--2800 $K$) with a few M7--M9 field dwarfs (2450-2650 $K$).  Thus, empirical comparisons favor HIP 71618 B as an M6--M8 low-to-intermediate-gravity dwarf with a temperature of $\approx$2600--2800 $K$ or less.

\subsection{Atmospheric Model Comparisons\label{sec:atmos}}

We use the solar metallicity BT-Settl model grid to further explore HIP 71618 B's atmosphere \citep{Allard2012}, covering temperatures and gravities of 400-4000 K and log(g) = 3--5.5, respectively, and adopting chemical abundances from \citet{Asplund2009}.    The models assume chemical equilibrium and include the effects of dust opacity and a cloud model.  Following \citet{Currie2023b}, we identify the radius for each model that minimizes the $\chi^2$ statistic for fitting our SCExAO/CHARIS JHK spectrum and the NIRC2 photometric point.  
We assume a distance of 58.2 pc when performing the comparison. 

\begin{figure}
\centering
\includegraphics[width=0.5\textwidth]{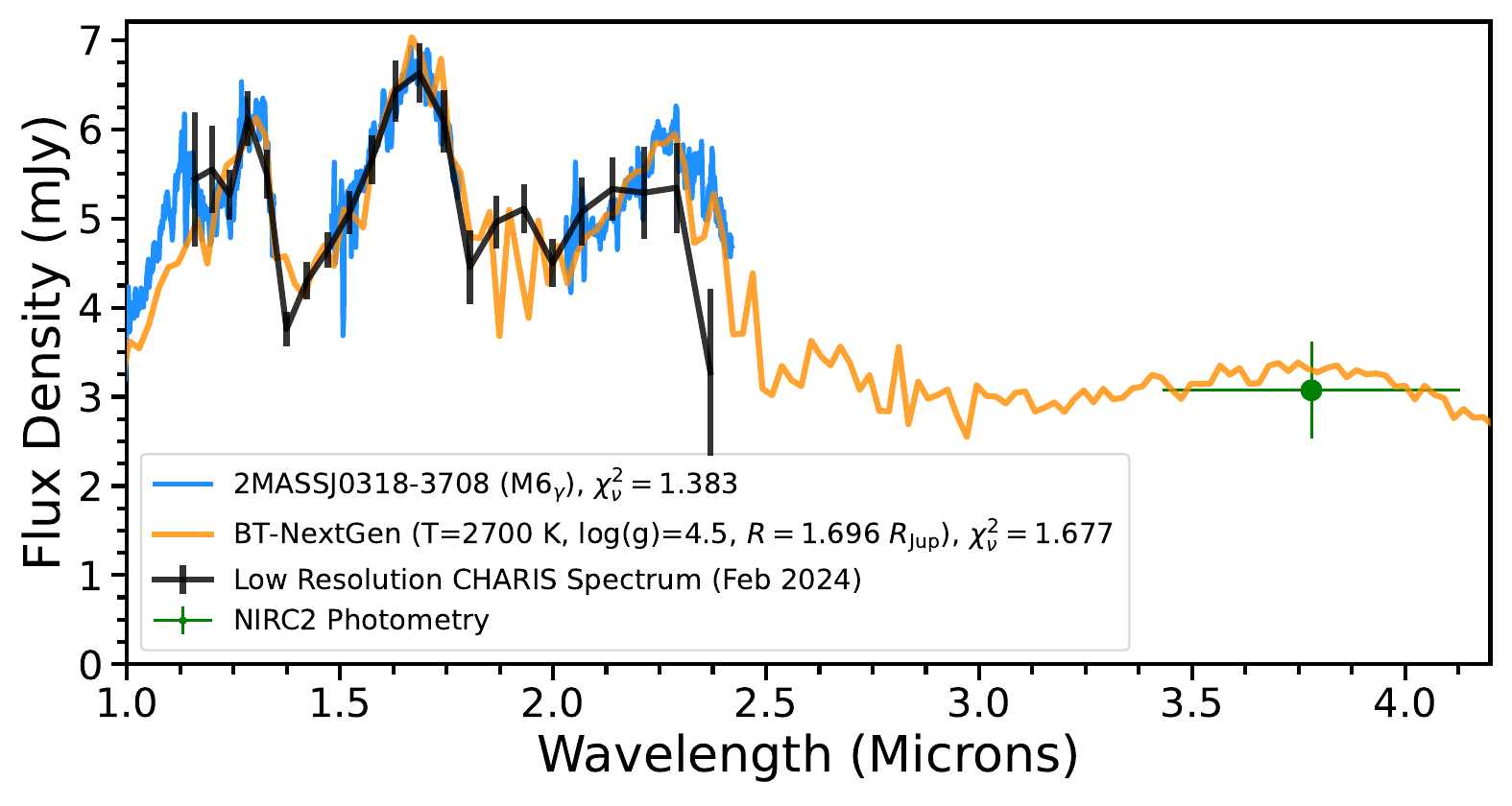}
\includegraphics[width=0.5\textwidth]{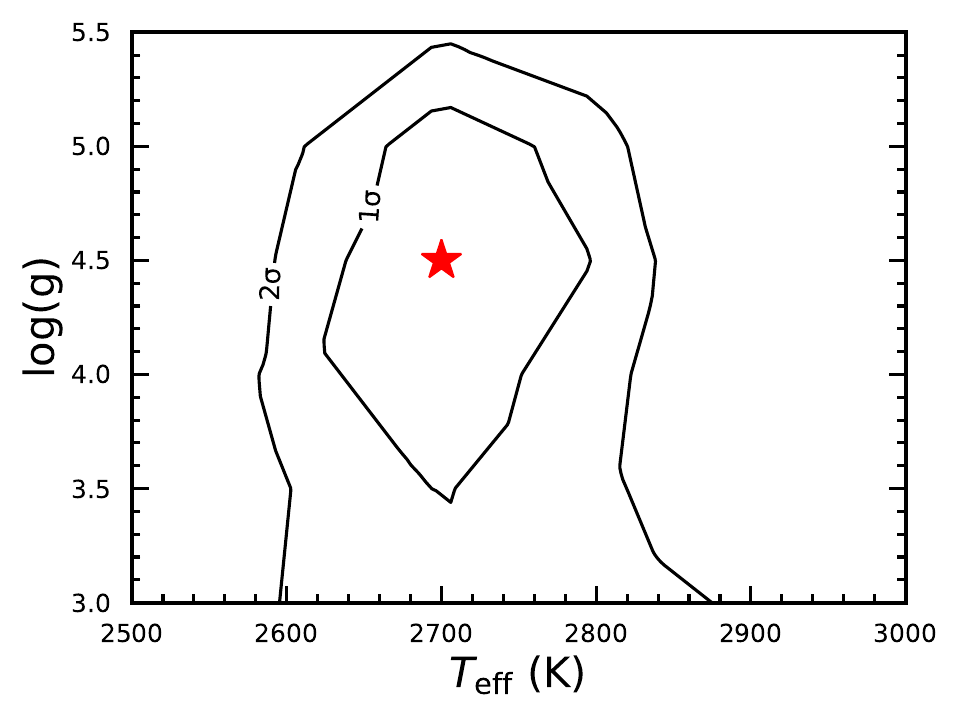}
   \caption{(Top) HIP 71818 b data compared to the best-fit model from the BT-Settl grids in the near-IR spectrum and NIRC2 datapoint.  (Bottom) Contour plot showing the confidence intervals for the BT-Settl model grid. The red star indicates the best-fit. }
   \label{fig:fit_btsettl} 
\end{figure}

The best-fitting BT-Settl model has a temperature of 2700 $K$, a gravity of log(g) = 4.5, and a radius of 1.696 $R_{\rm Jup}$ ($\chi^{2}_{\nu}$ = 1.677) (Figure \ref{fig:fit_btsettl}, top panel).
Figure \ref{fig:fit_btsettl}, bottom panel shows $\Delta\chi^2$ contours in the Teff–log(g) plane for the BT-Settl model fit.  While the surface gravity is poorly constrained, the 1$\sigma$ and 2$\sigma$ contours correspond to narrower ranges in temperatures of $\sim$2650--2800 $K$ and $\sim$2600--2850 $K$), respectively.

From the best-fitting BT-Settl model, we estimate HIP 71618 B's luminosity to be \( L = 5.54 \times 10^{23} \ \text{W} \), or \( \log\left(\frac{L}{L_\odot}\right) \approx -2.84 \).    
This luminosity is slightly higher than typical luminosities of old field M6--M8 dwarfs listed in \citet{Pecaut2013}, which have luminosities of \( \log\left(\frac{L}{L_\odot}\right)\) $\sim$ -2.98 to –3.28.  Adopting an age range of 20--200 Myr and assuming the COND evolutionary models to convert from luminosity to mass \citep{Baraffe2003}, HIP 71618 B's luminosity is consistent with a 30--100 $M_{\rm Jup}$ companion.

   \begin{deluxetable*}{lccr}[]
\tablecaption{MCMC Orbit Fitting Results}
\tablewidth{0pt}
  \tablehead{
  \colhead{Parameter} & \colhead{16/50/84\% quantiles} & Prior}
  \startdata
\multicolumn{3}{c}{Fitted Parameters (1,2)$^{a}$} \\ \hline
RV jitter (m/s)  &      ${0.14}_{-0.14}^{+57}$, ${0.10}_{-0.10}^{+53}$ & log-uniform \\
$M_{\rm pri}$ ($M_\odot$)   &      ${2.07}_{-0.24}^{+0.24}$, ${2.07}_{-0.25}^{+0.25}$  & Gaussian, $2.05 \pm 0.25$ \\
$M_{\rm sec}$ ($M_{\rm Jup}$)  &       ${60}_{-21}^{+27}$, ${65}_{-29}^{+54}$  & 65 $\pm$ 35 $M_{\rm Jup}$ (1-gaussian)$^{a}$, $1/M_{\rm sec}$ (2-flat/uniform)$^{a}$ \\
Semimajor axis $a$ (au)    &     ${11.1}_{-1.0}^{+1.3}$, ${11.3}_{-1.0}^{+1.3}$ & $1/a$  (log-uniform) \\
$\sqrt{e} \sin \omega$\tablenotemark{\rm *}    &     ${0.02}_{-0.43}^{+0.42}$, ${0.02}_{-0.45}^{+0.46}$ & uniform \\
$\sqrt{e} \cos \omega$\tablenotemark{\rm *}  &       ${0.73}_{-1.6}^{+0.18}$, ${-0.57}_{-0.31}^{+1.5}$  & uniform  \\
Inclination ($^\circ$)    &      ${98}_{-10}^{+20}$, ${97}_{-11}^{+18}$   &  $\sin i$ (geometric) \\
PA of the ascending node $\Omega$ ($^\circ$)  &      ${122.5}_{-6.2}^{+176}$, ${286}_{-169}^{+15}$ & uniform \\
Mean longitude at 2010.0 ($^\circ$) &        ${104}_{-77}^{+141}$, ${189}_{-149}^{+59}$ & uniform \\
Parallax (mas)   &      ${17.18}_{-0.12}^{+0.12}$, ${17.18}_{-0.12}^{+0.12}$  & Gaussian, 0 \\
\hline
\multicolumn{3}{c}{Derived Parameters(1,2)$^{a}$} \\ \hline
Period (yrs)     &    ${25.6}_{-3.1}^{+4.4}$, ${26.0}_{-3.2}^{+4.0}$ \\
Argument of periastron $\omega$ ($^\circ$)   &      ${176}_{-156}^{+164}$, ${178}_{-148}^{+151}$ \\
Eccentricity $e$    &   ${0.87}_{-0.22}^{+0.10}$, ${0.898}_{-0.22}^{+0.086}$   \\
Semimajor axis (mas)  &   ${191}_{-17}^{+23}$, ${193}_{-17}^{+22}$ \\
Periastron time $T_0$ (JD)    &    ${2463402}_{-7509}^{+1111}$, ${2463390}_{-7207}^{+1052}$\\
Mass ratio   &      ${0.028}_{-0.010}^{+0.013}$, ${0.030}_{-0.014}^{+0.027}$\\
\hline
\multicolumn{3}{c}{Predicted Locations (1,2)$^{a}$} \\ \hline
2027.0 & 0\farcs{}261 $\pm$ 0.009, 296.444$\pm$ 1.600$^{o}$, 0\farcs{}261 $\pm$ 0.009, 296.884$\pm$ 1.620$^{o}$ \\
2027.5 & 0\farcs{}247 $\pm$ 0.010, 296.044$\pm$ 2.047$^{o}$, 0\farcs{}248 $\pm$ 0.013, 296.914$\pm$ 2.286$^{o}$
\enddata
\tablenotetext{a}{Quantiles listed are for the simulations assuming 1) a gaussian companion mass prior centered on 65 $M_{\rm Jup}$ and 2) a flat prior. }
\label{tab:mcmc_result}
\end{deluxetable*}
\section{Dynamical Modeling\label{sec:astrometric_analysis}}

We constrained the dynamical mass of HIP 71618 B and its six Keplerian orbital parameters using the open-source Markov Chain Monte Carlo (MCMC) code \texttt{orvara} \citep{Brandt2021b}, combining absolute astrometry from HGCA \citep{Brandt2021b} with relative astrometry from SCExAO/CHARIS and Keck/NIRC2, as listed in Table \ref{astrom}.  Consistent with the primary's spectral type \citep[see][]{Pecaut2013}, we adopted a prior of $2.05 \pm 0.25~M_\odot$ for the host star and use the \textit{Gaia eDR3} parallax and uncertainties to define a prior on the parallax.

We carried out two separate fits, which differ in terms of the companion mass prior: (1) assuming a gaussian prior centered on 65 $M_{\rm Jup}$ with a standard deviation of 35 $M_{\rm Jup}$ and (2) a flat prior for the companion mass, as in \citet{Currie2023a}\footnote{Most of our prior work considered a log-uniform prior (\texttt{orvara}'s default option) and a flat prior \citep[e.g.][]{Currie2023b,Currie2025a,Bovie2025}.   However, adopting a true log-uniform prior causes a pile-up of solutions near unphysically low masses, where the prior has the highest probability density.  This numerical artifact is due to HIP 71618 B's relatively noisy astrometric acceleration measurement combined with a lack of evidence for orbital curvature.   HIP 99770 b has a similarly noisy acceleration measurement but does not suffer this artifact due to its more face-on orbit and larger angular displacement with time.}  The gaussian prior encompasses most of the luminosity-inferred mass range described in the previous section, with a slight bias against masses $<$30 $M_{\rm Jup}$, which are most consistent with ages less than 20 Myr.
Each \texttt{orvara} simulation consists of 100 walkers per temperature, 20 temperatures, 200,000 steps per walker, and thinned the chains by a factor of 50. The first 100 steps of the thinned chain are treated as burn-in.

\begin{figure*}
    \begin{flushright}
        \includegraphics[width=1.0\textwidth]{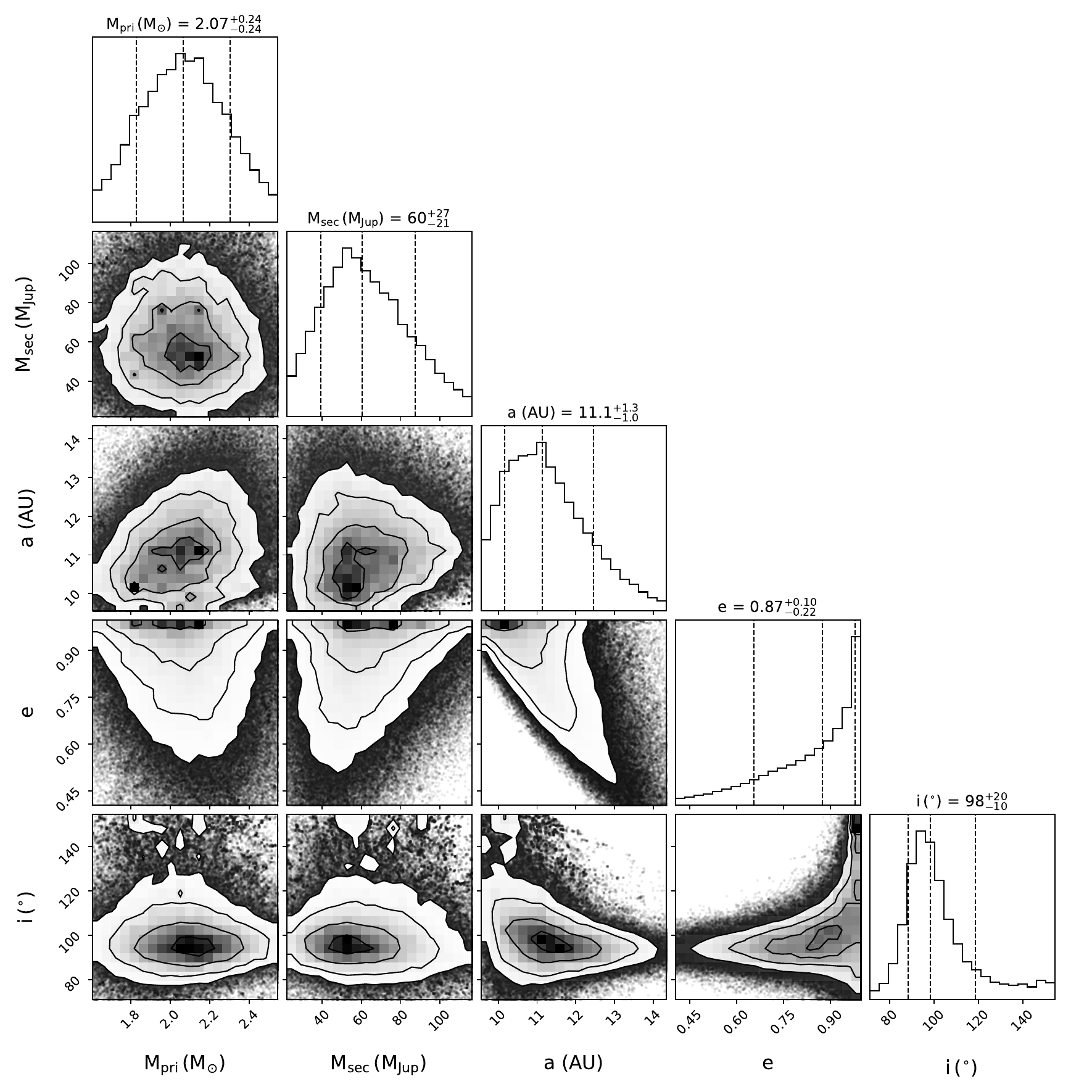} \\
    \vspace{-.95\textwidth}
    \includegraphics[width=0.44\textwidth]{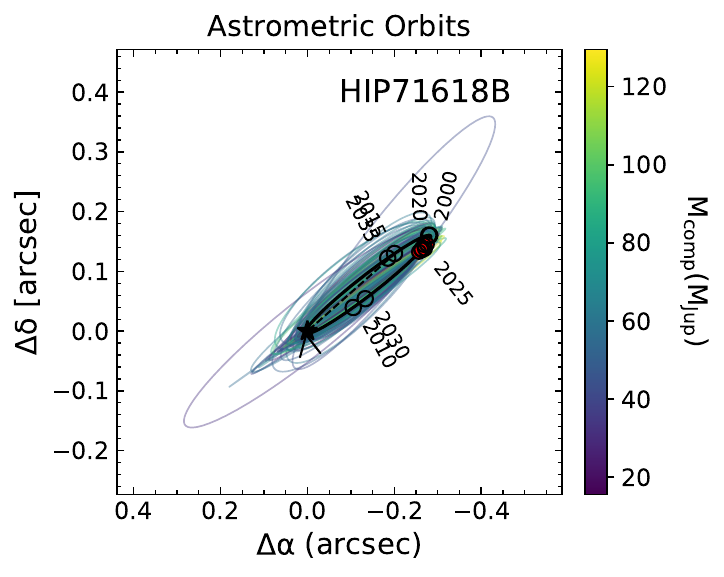} \\
    \vspace{0.575\textwidth}
    \end{flushright}
    \caption{Corner plot showing posterior distributions of selected orbital parameters from jointly modeling direct imaging and astrometric data assuming a gaussian companion mass prior centered on 65 $M_{\rm Jup}$.  The contour lines delineate the regions encompassing 68, 95, and 99\% of the posteriors.  The corner plot for the simulation adopting a flat companion mass prior is very similar except that its long tail of higher mass solutions results in a slightly larger range of masses within the 68\% confidence interval. }
    \vspace{-0.in}
    \label{fig:corner_plotgauss}
\end{figure*}

Table \ref{tab:mcmc_result} summarizes adopted priors for these two simulations and fitted parameters.  Figure \ref{fig:corner_plotgauss} 
displays the orbital parameter posterior distributions and an ensemble of
orbital solutions for the gaussian prior case.   Both simulations favor a semimajor axis of $\sim$ 11 au: ${11.2}_{-1.1}^{+1.4}$ and ${11.3}_{-1.2}^{+1.3}$ for the gaussian prior and flat companion mass prior simulations, respectively.  Both simulations also favor an orbit viewed nearly edge-on  (${96}_{-10}^{+18}$$^{o}$ and ${95}_{-12}^{+19}$$^{o}$). 

Our simulations find a broad eccentricity distribution favoring very high values (${0.87}_{-0.23}^{+0.11}$ for the gaussian prior, ${0.90}_{-0.22}^{+0.08}$ for the flat prior).  Other bound substellar companions likewise have broad eccentricity posteriors favoring high values, especially when a very small fraction of the companion's orbit is sampled \citep{Bowler2020,Blunt2023}.  Examining the posterior chains shows that HIP 71618 B's Hipparcos measurement likely favors a higher eccentricity, while other parameters show no preference.  To further test this hypothesis, we reran \texttt{orvara} with the HGCA astrometry turned off.  The resulting posterior distribution recovered our nominal results for primary mass, semimajor axis, and inclination, albeit with much larger errors.   However, it found less preference for a very high eccentricity ($e$ $\sim$ 0.67$^{+0.29}_{-0.41}$).

For both the gaussian and flat prior simulations, the median and nearly all of the 68\% confidence interval lie below the hydrogen limit \citep[$\approx$78.5 $M_{\rm Jup}$;][]{Chabrier2023} -- ${60}_{-21}^{+27}$ $M_{\rm Jup}$ and ${65}_{-29}^{+54}$ $M_{\rm Jup}$ --  thus favoring a brown dwarf interpretation for HIP 71618 B.   The corresponding mass ratio is low for a brown dwarf companion, just barely above the empirical turnover in mass ratios between brown dwarfs and planets identified from large ensembles of substellar objects ($q$ $\sim$  ${0.028}_{-0.010}^{+0.013}$ and $q$ $\sim$ ${0.030}_{-0.014}^{+0.027}$ for a gaussian and flat companion mass prior, respectively) \citep{Currie2023a,Currie2025a}\footnote{We find similar results for an \texttt{orvara} simulation using a log-uniform prior but removing artificially low companion mass solutions from the Markov chains: 46$^{+38}_{-22}$ $M_{\rm Jup}$.}.   

Given HIP 71618 B's luminosity and adopting the COND luminosity evolution models, the companion's dynamical mass is consistent with an object with an age between 15 Myr and 150 Myr (gaussian prior) or $>$15 Myr (flat prior), where HIP 71618 B's median posterior mass is most consistent with a $\approx$ 60-70 Myr-old object.  The COND models predict that an object with HIP 71618 B's luminosity lies below the hydrogen-burning limit if the age is $\lesssim$ 110 Myr -- i.e. if the age is roughly that of the Pleiades or less.

An age equal to or slightly younger than the Pleiades thus provides a self-consistent interpretation of HIP 71618 B's given its dynamical mass and luminosity, adopting the COND evolutionary model framework.   However, masses inferred from luminosity evolution models can show conflicts with precise dynamical masses \citep{Dupuy2014}; different luminosity evolution models can yield different mappings between luminosity, mass, and age \citep[e.g.][]{Brandt2021c}.

 \begin{figure}
\centering
   \includegraphics[width=0.4\textwidth,trim= 0mm 12mm 0mm 10mm,clip]{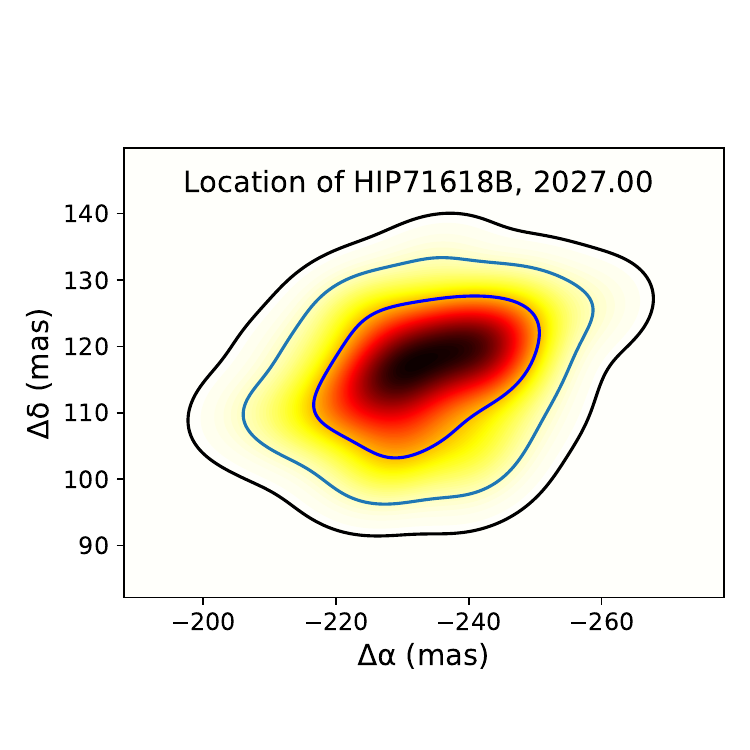}\\
    \includegraphics[width=0.46\textwidth,clip]{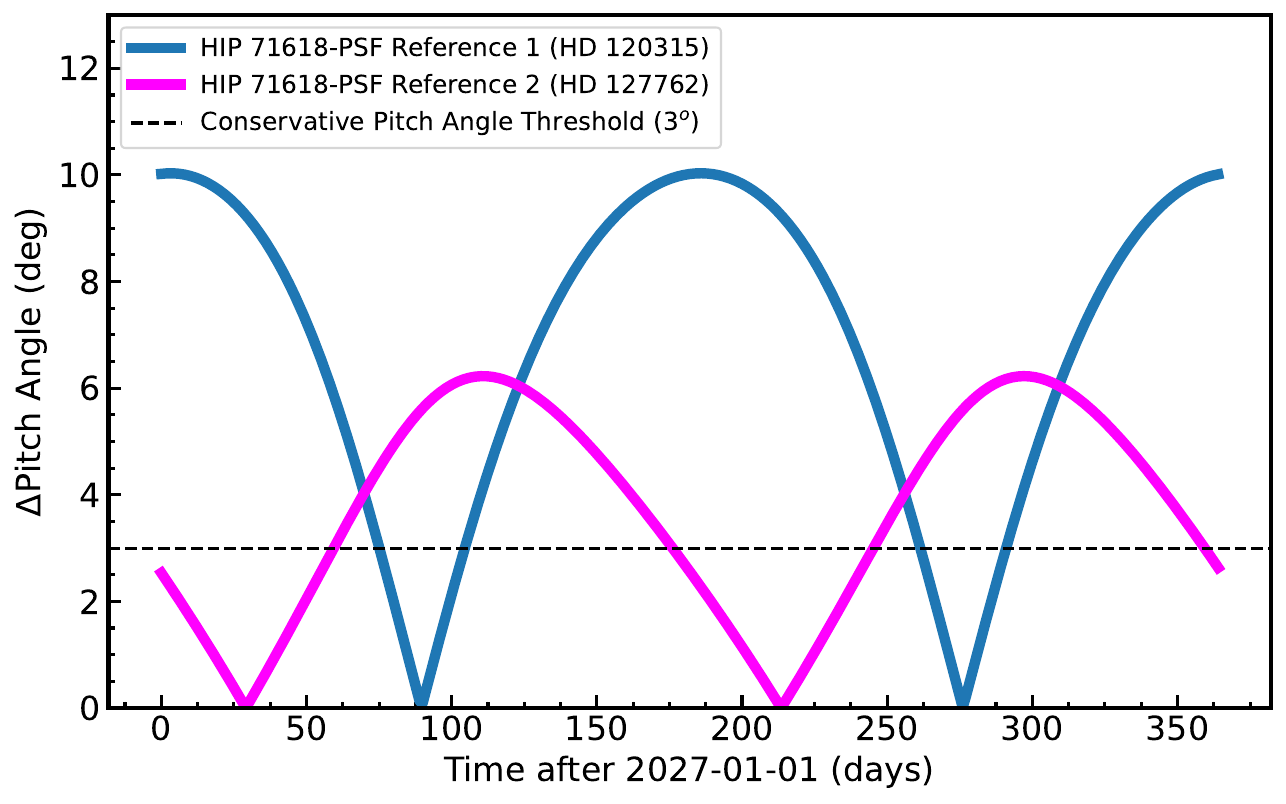}
   \caption{(top) 
   Predicted location of HIP 71618 B in 2027 January from the \texttt{orvara} dynamical modeling results presented in Table \ref{tab:mcmc_result}, assuming a gaussian companion mass prior.  (bottom) The pitch angle difference between HIP 71618 and two potential reference stars for dark-hole digging in 2027 with the conservative pitch angle threshold displayed as a dashed line.}
   \label{fig:cgidet} 
\end{figure}

\section{HIP 71618 B: A Target for the Roman Coronagraph Instrument Technology Demonstration\label{sec:cgi}}
HIP 71618 B is a $\sim$2600-2800 K companion to a bright primary (V $\sim$ 5.4) at a projected separation of $\approx$0\farcs{}3.   The Roman Coronagraph Instrument is predicted to yield deep ($\sim$10$^{-9}$--10$^{-8}$) contrasts at 575 nm over 0\farcs{}15--0\farcs{}45.   It has a single Threshold Technical Requirement (TTR5) for success\footnote{ \scriptsize{\url{https://roman.gsfc.nasa.gov/science/rsig/2021/Roman_Requirements_20201105.pdf}}}:
achieve a 5-$\sigma$ contrast better than     
10$^{-7}$ {at $\lambda_{\rm c}$ $\le$ 600 nm ($>$10\% bandpass) located 6--9 $\lambda$/D ($\sim$ 0\farcs{}3--0\farcs{}45) from a bright (V$_{\rm AB}$ $\le$ 5) star in under 10 hr\footnote{By construction, demonstrating the 10$^{-7}$ contrast threshold in more challenging configurations -- e.g. on a star slightly fainter than V = 5 or on a companion located between 3 and 6 $\lambda$/D -- would also fulfill TTR5.}.  The Roman Coronagraph passband available for TTR5 is centered on 575 nm.  

Here, we consider whether HIP 71618 
suitable for demonstrating TTR5.  We emphasize that this statement is \textit{not} equivalent to ``HIP 71618 is the \textit{only} system that can demonstrate TTR5".  Other suitable systems -- companionless stars or newly discovered systems with imaged companions -- assuredly exist.  However, they must be demonstrated from future analyses \textit{\underline{in peer-reviewed studies}} before they can be properly considered.

\subsection{HIP 71618 B Will Lie Within The Roman Coronagraph Dark Hole Region}
We first predict HIP 71618 B's position during the technology demonstration phase from our best-fitting \texttt{orvara} orbits.  We assume that this phase occurs during the first six months of full Roman Coronagraph operations in January -- July 2027, assuming a launch date of October 2026 and three months of commissioning time.   
During this timeframe, HIP 71618 B should lie at $\rho$ $\sim$ 0\farcs{}26 $\pm$ 0\farcs{}01 in January 2027 and 0\farcs{}25 $\pm$ 0\farcs{}01 in July 2027 ($\sim$5--5.2 $\lambda$/D) from the star 
(Figure \ref{fig:cgidet}, top panel), well within the 0\farcs{}15--0\farcs{}45 dark hole region where the latest full coronagraph simulation (OS 11) and TVAC tests predict an approximately flat 5$\sigma$ contrast floor of $\approx$ 2$\times$10$^{-8}$ or lower.   

\subsection{HIP 71618 Has Plausible PSF Reference Stars}
Roman Coronagraph observations of HIP 71618 require suitable reference stars for digging dark holes, which the Roman Coronagraph Community Participation Program (CPP) team is now vetting \citep{Wolff_2024_RomanCPP}.  From the CPP spreadsheet of candidate PSF reference stars\footnote{\url{https://docs.google.com/spreadsheets/d/1p5r0VmjBCjXU25daJl5oJOPoPh1V79nuESbnwmca0s0/edit?usp=sharing}}, we identified HD 120315 -- flagged as ``good?" through 2025 August 1 and as ``B" no later than 2025 September 1-- as a plausible reference.  It is located 10 degrees from HIP 71618\footnote{The deepest publicly available data for HD 120315 are Keck/NIRC2 data obtained in 2014.  
Through at least 2025 August 1, the Roman CPP team listed HD 120315 as "data reduction in progress".
We processed these data and did not identify a stellar companion that would preclude the star from being a suitable PSF reference.}.
Additionally, we used the \textit{SIMBAD} database to search for other very bright stars ($V$ $\lesssim$ 3--3.5)
close to HIP 71618 in the sky that are plausible PSF references despite not being on the CPP's spreadsheet.   This search identifies at least one additional candidate PSF reference star, HD 127762 (A7IV), which lies within 6$^{o}$ of HIP 71618, lacks a bright known companion from optical interferometric data, and has an angular diameter below the 2 mas threshhold for inclusion ($\theta_{diameter}$ $\sim$ 1.77 mas) \citep{Baines2023}.   It was likely missed due to its \textit{Simbad}-listed V band magnitude being just 0.02 mag fainter than the nominal CPP cutoff ($V$ = 3.02).  

The instrument's thermal stability limits its performance. Differences in the pitch angle greater than 3$^{o}$ between the telescope pointing to a target vs. PSF reference and the Sun may degrade contrasts.   However, for over 90\% of a calendar year, HIP 71618's pitch angle is within 3$^{o}$ of at least one of the two aforementioned PSF stars (Fig. \ref{fig:cgidet}, bottom panel).  

A TTR5-achieving observation of HIP 71618 can be efficiently scheduled.  HIP 71618 and HD 102315 both have ecliptic longitudes $>$ 54$^{o}$ and are thus in or very close to the Roman Continuous Viewing Zone \citep[CVZ,][]{Rose2023}, enabling highly flexible scheduling, while HD 127762 is very close to the CVZ.  Keepout maps generated for HIP 71618 and its two candidate PSF references\footnote{These can be easily modifiable from \url{https://github.com/roman-corgi/roman_pointing/blob/main/Notebooks/roman_pointing_demo.ipynb}.} identifying non-scheduleable epochs show that HIP 71618 and HD 102315 can be imaged by Roman year-round in 2027, while HD 127762 is observable except for two $\sim$ 40-day blocks (70--110 and 260-310 days after January 1).

\subsection{A High Signal-to-Noise Ratio Detection Of HIP 71618 B With The Roman Coronagraph Would Fulfill The Instrument's Core Technology Demonstration Goal}
To estimate HIP 71618 B's contrast at 575 nm and other passbands, we use predicted contrasts from the \citeauthor{LacyBurrows2020} and BT-Settl/BT-NextGen atmosphere models and compare these results to empirical spectra and and photometry for field dwarfs and younger, lower gravity dwarfs.   
For the \citeauthor{LacyBurrows2020} models, we focus on temperatures of 2600--2800 $K$ and a gravities of log(g)=4--5 and the radii for each temperature/gravity combination that minimizes the $\chi^{2}$ value.  These models incorporate new line lists particularly relevant for the Roman Coronagraph passbands \citep[see ][]{Currie2023a,Bovie2025}.  For the older BT-Settl/NextGen models, we focus on the best-fitting model phase space (temperature, gravity, radius) found in Section \ref{sec:atmos}.

For an empirical calibration of HIP 71618 B's expected contrast, we add optical spectra for young, low-gravity members of the TW Hya and the Tucana-Horologium moving groups ($\sim$ 10 Myr and 45 Myr old, respectively) drawn from \citet{Luhman2017} and \citet{Luhman2023,Luhman2024}:  	
2MASS J00514561-6227073/Gaia DR3 4710164571939123456 (M5.5-M6), 2MASS J00200551-5359372/Gaia DR3 4923392999353596288 (M7.25), and TWA 26 (M8.5).   The spectra do not extend to the near-IR filters where we detect HIP 71618 B.  Thus, we normalize these spectra within a 100 nm bandpass centered on 850 nm to the best-matching subset of BT-Settl/NextGen models that fit also HIP 71618 B within $\chi^{2}_{\rm \nu}$ $<$ 2.    The 700-1000 nm portion of the M5.5-M6, M7.25, and M8.5 standard spectra are best matched by BT-NextGen models of 2800 $K$/log(g) = 4.5, 2700 $K$/log(g) = 4.5, and 2600 $K$/log(g) = 4.5, respectively.


Finally, we add measurements of HD 1160 B obtained with SCExAO/VAMPIRES \citep{Norris2015,Lucas2024}.   HD 1160 B has a spectral type similar to or slightly earlier than HIP 71618 B, likely has a similar mass, and likely has a comparable age but orbits its host star at a far wider separation making its detection significantly less challenging \citep{Nielsen2012,Garcia2016}.  We observed HD 1160 with SCExAO and the upgraded VAMPIRES detector on 2023 July 11 in excellent seeing conditions in four passbands centered on 614 nm, 670 nm, 720 nm, and 760 nm. The observing sequence covers 17 minutes of integration time and 14 degrees of parallactic angle rotation ($\sim$ 10-12 $\lambda$/D at HD 1160 B's separation).   The data were reduced using the VAMPIRES Data Processing Pipeline \citep{Lucas2024}; we performed PSF subtraction using the GreeDS algorithm \citep{Pairet2021}, an improvement upon the widely-used Principal Component Analysis-based subtraction approach \citep{Amara2012,Soummer2012-KLIP}.  This approach yielded high SNR detections at 720 and 760 ($\sim$ 29$\sigma$ and 34$\sigma$), a weaker detection at 670 nm ($\sim$ 6.3$\sigma$), and a 2.4-$\sigma$ residual at 614 nm consistent with a noisy detection.   The throughput-corrected contrasts at 610 nm, 670 nm, 720 nm, and 760 nm are then 1.3$\times$10$^{-6}$, 3.8$\times$10$^{-6}$, 1.7$\times$10$^{-5}$, and 3.0$\times$10$^{-5}$, respectively.   Around HIP 71618, HD 1160 B would have a contrast $\sim$14\% fainter in all bands.

 \begin{figure}
\centering
        \includegraphics[width=0.487\textwidth,clip]{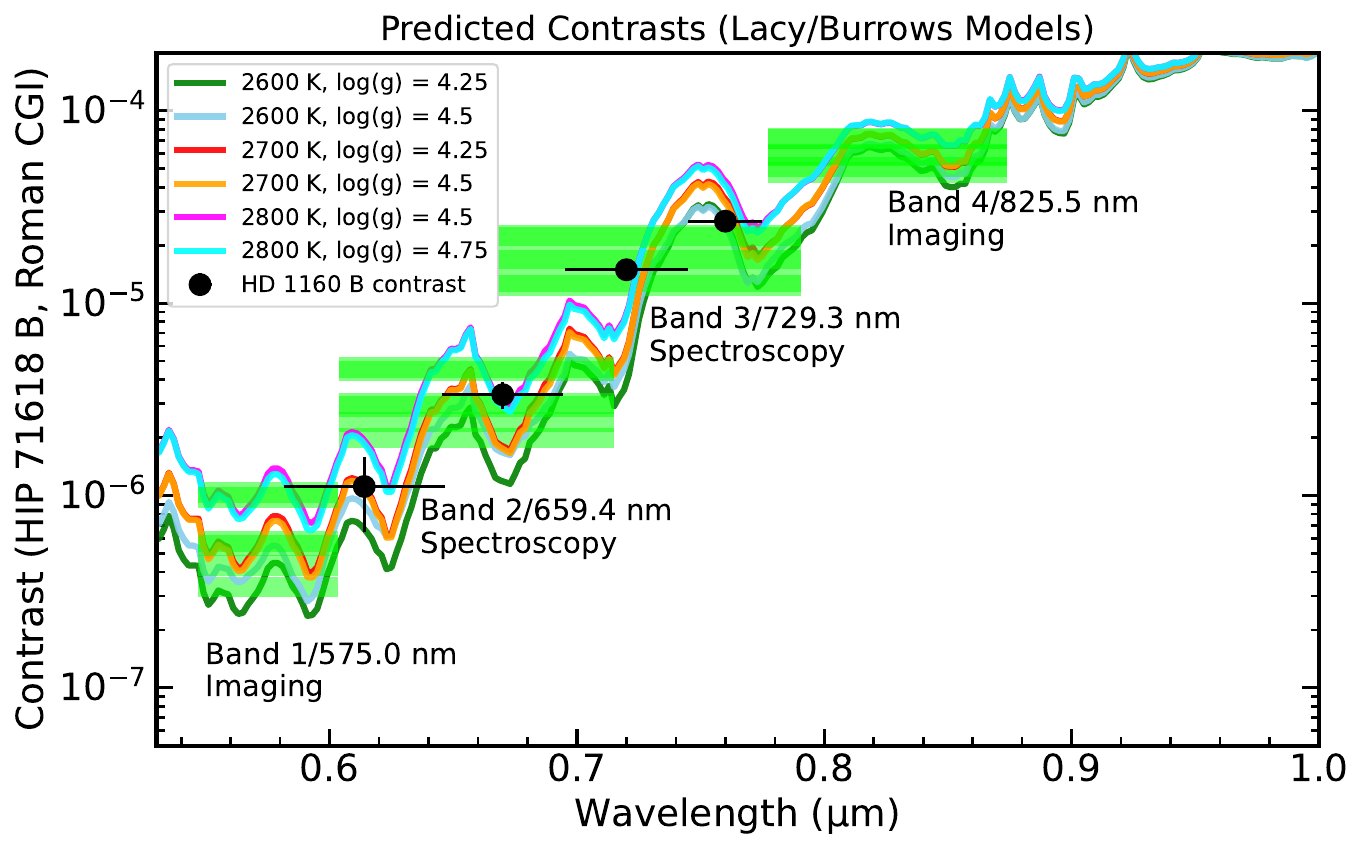}\\
    \includegraphics[width=0.487\textwidth,clip]{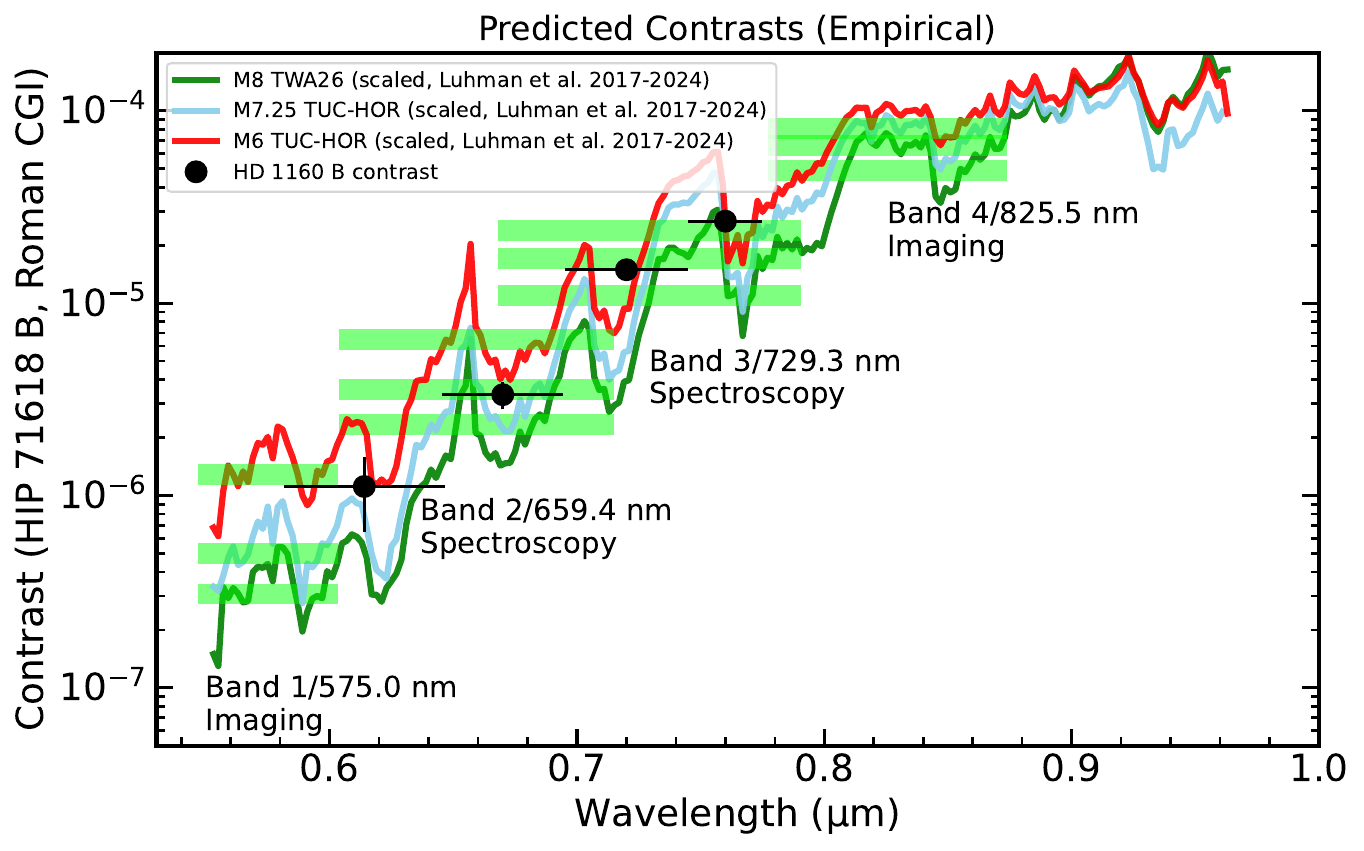}
   \caption{(Top) 
  (top) Predicted contrast for HIP 71618 B in the coronagraph's passbands for different atmosphere models from the updated \citet{LacyBurrows2020} grid distribution.   (bottom) Predicted contrasts for HIP 71618 B drawn from flux-normalized optical spectra from the young/low-gravity spectral standards presented in \citet{Luhman2017} and \citet{Luhman2023,Luhman2024}.  The black data points show contrasts for HD 1160 B -- a substellar companion with near-IR colors, a mass, and age similar to HIP 71618 B --  if it orbited HIP 71618.} 
   \label{fig:cgidet2} 
\end{figure}



Figures \ref{fig:cgidet2} shows our results. 
The Lacy/Burrows models predict 575 nm contrasts of $\sim$5.7$\times$10$^{-7}$ from the best-fit temperature and gravity (2700 K, log(g) = 4.5) and 3--10$\times$10$^{-7}$ for other models displayed (top panel).   The predicted contrasts agree very well with the measured optical contrasts for HD 1160 B (black dots).  Thus, assuming these models accurately predict HIP 71618 B's 575 nm brightness, a SNR = 28.5 detection of HIP 71618 B will likely demonstrate a 5-$\sigma$ contrast of 10$^{-7}$ contrast.  A SNR = 50 detection would demonstrate this contrast for the upper range of 575 nm contrast predictions.

Contrasts inferred from the optical spectra of young brown dwarfs largely agree with the Lacy/Burrows model predictions (bottom panel).  Using the M6 Tuc-Hor standard to guide our estimates, HIP 71618 B should have a contrast of $\sim$ 1.3$\times$10$^{-6}$.  The M7.25 and M8.5 standards imply steeper contrasts of $\sim$ 5.0$\times$10$^{-7}$ and 3.1$\times$10$^{-7}$, respectively: in agreement with model predictions.  HD 1160 B's contrasts appear to be in strongest agreement with those from the M7.25 spectral standard.  Adopting 575 nm contrast estimates drawn from these standards, a SNR = 16--65 detection of HIP 71618 B would demonstrate a 5-$\sigma$ contrast of 10$^{-7}$ contrast\footnote{We explored alternate spectral standards drawn from \citet{Luhman2017} and \citet{Luhman2023,Luhman2024}, especially in the M6--M7 range, but found nearly identical results.  While the BT-Settl/NextGen models well reproduce the colors and spectral features of our spectral standards in the red optical, they appear to overpredict the standards' flux densities shortwards of 700 nm.  In contrast, they provide excellent fits to spectral of old M dwarfs from the MELCHIORS library \citep{Melchiors2024}, hinting at challenges with fitting optical spectra of M dwarfs at low gravities.}.

To estimate the exposure time needed to carry out TTR5-fulfilling coronagraph observations of HIP 71618, we use the Roman Coronagraph Instrument Exposure Time Calculator (Corgi-ETC)\footnote{\url{https://github.com/roman-corgi/corgietc}}.   To achieve 10$^{-7}$ 5-$\sigma$ contrasts for HIP 71618 in the optimistic and conservative performance scenarios, Corgi-ETC predicts on-source observing times of 1.5 hrs and 1.75 hrs, respectively, which would yield an SNR = 25 detection of HIP 71618 B for a contrast of 5$\times$10$^{-7}$.   For the worst-case scenario (a 575 nm contrast of = 3$\times$10$^{-7}$), these times only slightly increase (1.9 hrs and 2.5 hrs). Assuming the coronagraph primer observing efficiency estimate (24 hours clock time for 14 hours observing time)\footnote{\url{https://roman.ipac.caltech.edu/docs/RomanCoronagraphPrimer_Current.pdf}},  HIP 71618 observations require nominal clock time of 2.6 hrs and 3 hours for the optimistic and conservative scenarios.  Worst-case scenario values are 3.3 hrs and 4.3 hrs.  All of these values are well below the time limits specified for TTR5 (10 hours).

Thus, current evidence shows that HIP 71618 is a highly suitable target for demonstrating TTR5 due to its predicted contrast, location, primary brightness, availability of (candidate) PSF reference stars, and ease of scheduling.    Follow-up deep high-contrast imaging observations of both stars could confirm the suitability of PSF reference stars.

\section{Summary}
Our study further demonstrates the utility of programs like OASIS that combine direct imaging and precision astrometry to discover substellar companions.   HIP 71618 B likely is a moderate mass brown dwarf ($\approx$60 $M_{\rm J}$) at 11 au but with a mass ratio of $q$ $\approx$ 0.03, placing it within the brown dwarf desert where both indirect methods like radial-velocity and direct imaging infrequently find objects \citep[e.g.][]{Nielsen2019,Stevenson2023,Currie2025a}.   Current data favor a very high eccentricity for HIP 71618 B compared to many other substellar companions detected by imaging and astrometry \citep[e.g. HD 33632 Ab][]{Currie2020a,ElMorsy2024b}, although the eccentricity posterior distribution is quite broad.  Early orbital modeling for some substellar objects with relative astrometry covering a short time baseline likewise favored very high eccentricities: subsequent monitoring favored lower eccentricities \citep[e.g.][]{Blunt2023}.  Further astrometric monitoring and relative RV measurements will enable far better eccentricity constraints, as has been found for HIP 99770 b \citep{Bovie2025,Winterhalder2025}.

HIP 71618 B could play a unique role in the success of the Roman Coronagraph by being a first target on which to achieve TTR5.     HIP 71618 is bright enough (V = 5.39) that the instrument should lose little of its nominal performance (benchmarked at V $\le$ 5). The location of HIP 71618 and potential PSF reference stars within the Roman CVZ significantly relaxes scheduling constraints compared to targets at lower galactic latitudes.  A high SNR detection of HIP 71618 B at 575 nm would fulfill TTR5, the instrument's sole success criterion.

HIP 71618 is especially important because current alternatives for demonstrating TTR5 are, at best, problematic.  For example, achieving TTR5 ``by analysis" of the residual noise around a companionless star near a PSF reference is possible in principle.  However, in this approach, both the PSF reference stars \textit{and} suitable ``companionless" star targets must be vetted (e.g. for contaminating binaries at the $\gtrsim$10$^{-5}$ level, D. Savransky 2025, pvt. comm.) prior to scheduling consideration.   As of October 24 2025 ($\sim$ 11-18 months before launch), there are no publicly announced companionless star vetting programs within the CPP.

A far more striking alternative is to demonstrate TTR5 by reimaging a known, cool substellar companion found from IR high-contrast imaging.
However, as has been pointed out many times before \citep[e.g.][]{ElMorsy2024a}, the peer-reviewed literature currently lacks any known imaged companion whose redetection with Band 1 is 1) demonstrated to be likely and 2) would demonstrably fulfill TTR5.  E.g. $\beta$ Pic b is undetectable at 575 nm due to the system's debris disk.  Published models \citep{LacyBurrows2020} suggest that 51 Eri b is too faint. Updates to these models likewise show HR 8799 e, HD 206893 B, and HIP 99770 b are likely too faint ($\sim$10$^{-9}$--10$^{-11}$ contrast) \citep[B. Lacy, pvt. comm.;][]{Bovie2025} \footnote{A focus on mature, radial-velocity (RV) detected Jupiter-mass planets only exacerbates target problems.
Almost all well characterized, RV-detected mature planets around V $<$ 5 stars lie interior to 3--6 $\lambda$/D for Roman at 575 nm; others require steep contrasts $\lesssim$10$^{-9}$ for detection (e.g. $\upsilon$ And): only feasible if its current best laboratory contrasts are realized in flight and 100 times more challenging than TTR5's 10$^{-7}$ contrast threshhold.  The majority also lack strong orbital constraints in the Exoplanet Imaging Database (due to lack of inclination estimates; resulting uncertainties in time of periapsis passage) (D. Savransky, 2025, pvt. comm.).  They are not currently considered to be the most straightforward targets for companion detection  (D. Savransky, 2025, pvt. comm.).}.  As of October 24 2025, there a dedicated public-facing companion target search within the context of the Roman CPP program, nor is there elsewhere peer-reviewed work from the CPP team clearly demonstrating a sample of suitable imaged companions besides HIP 71618 B..

The suitability of HIP 71618 as a technology demonstration target should motivate timely, deep vetting of reference stars to confirm their lack of binarity at deeper contrasts.
Note that HIP 71618 is $\sim$30-40\% fainter than the nominal target brightness benchmark of V = 5, and HIP 71618 B will be located slightly interior to the 6--9 $\lambda$/D scoring region but within the region (3--9 $\lambda$/D) that both simulations and TVAC tests demonstrate that the coronagraph will yield a deep-contrast dark hole\footnote{See demonstrated here \url{https://workshop.ipac.caltech.edu/romancgi24/talks/Day1_Poberezhskiy.pdf} (page 10, comparing the 3--9 $\lambda$/D and 6--9 $\lambda$/D performance) and supported here \url{https://roman.ipac.caltech.edu/page/cgi-contrast-curves-html}}.  Achieving TTR5 on HIP 71618 will therefore present a slightly conservative assessment of the actual instrument performance.

Finally, HIP 71618 B is also well-suited for commissioning observations prior to a TTR5 demonstration and for different purposes.  Its modest predicted 730 nm contrast (10$^{-5}$; see Figure \ref{fig:cgidet}, bottom left panel) suggests it is well suited for long-slit spectroscopy, even if the instrument is not yet tuned well enough to achieve TTR5.  Thus, in addition to satisfying TTR5, the system may be well suited for  an early demonstration of the instrument's ability to derive atmospheric properties of substellar objects.

\begin{acknowledgments}
\indent We thank the anonymous referee for timely and helpful suggestions that improved the quality of this paper and Dimitry Savransky for discussions about current Roman CPP target selection preparations.  We are extremely grateful to Kevin Luhman for helpful guidance on low-gravity spectral standards and for sharing the TWA 26 spectrum.\\
\indent The authors acknowledge the very significant cultural role and reverence that the summit of Mauna Kea holds within the Hawaiian community.  We are most fortunate to have the opportunity to conduct observations from this mountain.   \\
\indent This work is generously supported by National Science Foundation (NSF) Astronomy and Astrophysics Grant \#2408647 and NASA-Keck Strategic Mission Support Proposal.  We are grateful for the continued valuable work and expert guidance of NSF and NASA-Keck personnel, especially in challenging times. M.K. is supported by JSPS KAKENHI (grant No. 24K07108).  A.Z. acknowledges support from ANID -- Millennium Science Initiative Program -- Center Code NCN2024\_001 and Fondecyt Regular grant number 1250249.\\
\indent The development of SCExAO was supported by JSPS (Grant-in-Aid for Research \#23340051, \#26220704 \& \#23103002), Astrobiology Center of NINS, Japan, the Mt Cuba Foundation, and the director's contingency fund at Subaru Telescope.  CHARIS was developed under the support by the Grant-in-Aid for Scientific Research on Innovative Areas \#2302.  SCExAO’s adaptive optics loops and high-speed data acquisition are handled by  the CACAO package, which is supported by NSF award 2410616.  Some of the data presented herein were obtained at the W. M. Keck Observatory, which is operated as a scientific partnership among the California Institute of Technology, the University of California and the National Aeronautics and Space Administration. The Observatory was made possible by the generous financial support of the W. M. Keck Foundation. Ziying Gu acknowledges the support from the Forefront Physics and Mathematics Program to Drive Transformation (FoPM), a World-leading Innovative Graduate Study (WINGS) Program at the University of Tokyo.

\end{acknowledgments}

\software{
          corgi-etc:  Roman Coronagraph Community Participation Program (CPP)\footnote{\url{https://github.com/roman-corgi/corgietc}},
          roman$\_$pointing:  Roman Coronagraph Community Participation Program (CPP)\footnote{\url{https://github.com/roman-corgi/roman_pointing/}}. 
          }

\bibliography{bibliography}{}
\bibliographystyle{aasjournal}

\end{document}